\documentclass[twocolumn,showpacs,pre,aps,amssymb,amsmath]{revtex4}
\usepackage{graphicx}
\DeclareMathSymbol{\openbra}{\mathord}{symbols}{"68}
\DeclareMathSymbol{\closeket}{\mathord}{symbols}{"69}

\newcommand{\bd}{\begin{displaystyle}}
\newcommand{\ed}{\end{displaystyle}}
\begin{document}
\title{Fractal atom-photon dynamics in a cavity

{\it To the memory of my beloved daughter Maria}}

\author{S.V. Prants}
\affiliation{Laboratory of Nonlinear Dynamical Systems,
V.I.Il'ichev Pacific Oceanological Institute of the Russian Academy of Sciences,
690041 Vladivostok, Russia}

\begin{abstract}
Nonlinear dynamics in the fundamental interaction between a two-level 
atom with recoil and a quantized radiation field in a high-quality cavity is studied. 
We consider the strongly coupled atom-field system as a quantum-classical 
hybrid with dynamically coupled quantum and classical degrees of freedom.
We show that, even in the absence of any other 
interaction with environment, the interaction of the purely quantum 
atom-field system with the external atomic degree of freedom provides 
the emergence of classical dynamical chaos from quantum electrodynamics. 
Atomic fractals with self-similar intermittency of 
smooth and unresolved structures are found in the exit-time scattering 
function.
Tiny interplay between all the degrees of freedom is responsible for 
dynamical trapping of atoms even in a very short microcavity. 
Gedanken experiments are proposed to detect manifestations of atomic fractals 
in cavity quantum electrodynamics.
\end{abstract}
\maketitle
\section{Introduction}

The emergence of classical dynamical chaos from more profound quantum 
mechanics is one of the most intriguing problems in physics 
\cite{Chir79,Z81,Haake,Gutz}. Dynamical chaos in classical mechanics is 
a special kind of random motion in dynamical 
systems without any noise and random parameters that is characterized by 
sensitive dependence on initial conditions in a bounded phase space. 
The current consensus is that {\it isolated bounded quantum systems} do not 
show sensitive dependence on initial conditions in the same way as 
classical systems because their evolution is unitary. So the question is: 
What is the fundamental mechanism of arising classical chaos from 
quantum mechanics? 

The purpose of this lecture is to show how classical Hamiltonian chaos with 
sensitive dependence on initial conditions, positive Lyapunov exponents, 
fractal properties of underlying phase space, and anomalous statistical 
characteristics may arise from quantum electrodynamics of a single atom 
strongly interacting with a quantized mode of the electromagnetic radiation 
field in a high-quality cavity. The study of the fundamental atom-photon 
interaction constitutes the rapidly growing field of cavity quantum 
electrodynamics (for a review see \cite{QED,W92,RBN,VWW}). The experimental state 
of the art has reached in this field the stage where the transition from 
classical to quantum dynamical regimes can now be probed directly. Atoms  
and photons, confined in a high-quality cavity, are ideal objects to study 
quantum-classical correspondence and quantum chaos. 

Real quantum systems are not isolated. They interact with their environment
and in attempts to measure their states  with classical measuring devices 
which should, by virtue of their purpose, be in unstable states. So we are 
dealing with quantum-classical hybrids.  It is not a 
simple question which degrees of freedom should be considered in a given 
physical situation as  quantum ones and which ones as classical. 
Let us consider an excited atom 
in a single-mode high-quality cavity whose frequency is close to the frequency 
of one of the atomic electrodipole transitions. The atom emits a photon in the 
cavity mode and goes to a lower-lying state. Under the conditions of the  
strong atom-field coupling, the photon may be reemitted and reabsorbed by the atom 
many times. Sooner or later, the atom will emit a photon in one of the other modes 
of the electromagnetic field that are not sustained by the cavity (for example, 
transverse modes in an open Fabry-Perot cavity) and the photon will be lost in 
a surrounding, the process known as spontaneous emission. In real experiments, 
the cavity relaxation due to the losses in the cavity walls cannot be neglected. 
It is usually modeled by a weak coupling of the selected cavity mode with a bath 
of harmonic oscillators (spanning a wide frequency range)   
in thermal equilibrium at a given temperature. Sooner or later, the atom will relax 
to  a lower-lying state that is not resonant with the cavity mode. It seems to be 
reasonable to treat two near-resonant atomic levels, strongly coupled to  
a cavity mode, as a quantum dynamical system weakly coupled to the other electromagnetic 
modes, the other atomic levels and the cavity walls. An infinite number of the 
respective degrees of freedom, which are coupled to the quantum dynamical system's 
degrees of freedom but are not affected by them, forms an inexhaustible  external 
reservoir (that may be treated classically) usually called ''the environment". 
It is a kind of the external coupling which is inevitably present in reality. 

There is another kind of coupling called the dynamical coupling. For example, 
if the average number of photons in the cavity mode is sufficiently large, one may 
treat the field as a classical object dynamically coupled to quantized atoms 
(the semiclassical approximation) and take into account the feedback effect 
of the atoms on the radiation field. The semiclassical approximation breaks down 
when we deal with single cold atoms and photons in a cavity. In this situation we should 
take into account the translational (external) atomic degree of freedom. 
When a cold atom 
emits and absorbs photons its momentum  and position may vary significally due to 
the recoil effect. The external atomic degree of freedom, which may be treated 
classically if the values of
the atomic momentum are greater than the photon momentum, is dynamically coupled to 
the internal atomic and field degrees of freedom which are treated quantum 
mechanically. We show in this lecture that, even in the absence of any other 
interaction with the environment, the interaction of the quantum atom-field system 
with the external atomic degree of freedom provides, under appropriate conditions,
the emergence of classical 
chaos from quantum electrodynamics.     

\section{Early studies of classical chaos in the atom-field interaction}

The discovery that a single-mode laser, a symbol of coherence and stability,
may exhibit deterministic chaos is especially important not only because
lasers are one of the main instruments in physics but lasers, as well,
provide almost ideal systems to test general ideas in quantum mechanics and
statistical physics. From the standpoint of nonlinear dynamics, laser is an
open dissipative system that transforms an external excitation into a
coherent output in the presence of loss. Some manifestations of strange
attractors and dissipative chaos have been observed with different types of
lasers (for a review, see \cite{Mil,Har,Kh}). A collection of identical two-level
atoms, interacting with a single-mode electromagnetic field, provides the
simplest model for laser dynamics. Because of a large number of atoms (and
photons), laser dynamics can be adequately described in the so-called
semiclassical approximation, where one treats atoms as two-level objects,
which may be described by the Bloch equations for two components of the
collective electrodipole polarisation and collective atomic population
inversion, interacting with a field mode to be governed by the classical Maxwell
equations for the field strengths with the right hands depending on the
atomic polarisation. With a single-mode homogeneously broadened laser,
operating in resonance with the gain center, these five equations can be
reduced to three real-valued equations for slowly varying amplitudes which
have been shown to be equivalent to the well-known Lorentz model for fluid
convection \cite{Hak}.

Practically in the same time, ideas of dynamical chaos have been explored 
with fundamental models of the matter-radiation interaction, comprising of a
collection of two-level atoms {\it interacting with their own radiation
field} in a perfect single-mode cavity without any loss and external
excitation. It has been shown theoretically and numerically \cite{Bel} that
the semiclassical Maxwell--Bloch equations, following from the Dicke
Hamiltonian \cite{Dicke}, may demonstrate {\it Hamiltonian semiclassical
chaos} if one goes beyond the so-called rotating-wave approximation, i.~e. if
one takes into account energy non-conserving terms in the Hamiltonian (for
details, see the beginning of the next section). This mechanism of arising
Hamiltonian semiclassical chaos is rather weak under realistic assumptions.
On the other hand, rotating-wave approximation suppresses chaos with atoms at
rest due to existence of an additional integral of motion, the conserving
interaction energy. It has been shown in a series of our papers
\cite{JETPL97,PLA97,JETP99,PRE99,PRE00} that Hamiltonian chaos may arise
within the rotating-wave approximation under conditions of a modulation of
the atom-field interaction. In a natural way it occurs when atoms move
through a cavity in a direction along which the cavity sustains a
standing-wave field that is periodic in space \cite{PLA97,PRE99}. When moving
with a constant velocity, atoms ``see'' the field whose strength changes in
time periodically. It breaks down the interaction-energy integral and may
cause Hamiltonian semiclassical chaos of a homoclinic type
\cite{PLA97,PRE99,PRE00}. There are another ways of modulations to be
considered in \cite{JETP99,IKP}. The modulation of the detuning between the atomic
transition frequency and the frequency of the field mode has been shown to
produce parametric instability and Hamiltonian semiclassical chaos
\cite{JETP99}. Structural Hamiltonian chaos \cite{IKP} may arise as a result
of a harmonic modulation of cavity length which causes the respective
oscillations of the nodes of a standing-wave.

Generally speaking, the model of the atom-field interaction should involve
not only the internal atomic and field degrees of freedom but also the
center-of-mass motion of the atom. When emitting and absorbing photons, atoms
not only change their internal states but their position and momentum are
changed as well due the photon recoil effect. This effect may be neglected if
one deals with Rydberg atoms interacting with a microwave field in a cavity
(as it has been done in \cite{JETPL97,PLA97,PRE99,PRE00,IKP}) or with thermal usual atoms interacting with a visible light. It has been
theoretically and numerically shown in \cite{PK01,PS01} that Hamiltonian
semiclassical chaos may arise with cold atoms with recoil in a standing-wave
microcavity.

In trying to describe adequately dynamics of single atoms and photons in a
high-quality cavity, one should go beyond the semiclassical approximation and
treats the atom-photon interaction on a quantum ground. The fully quantum
model of the interaction between a single two-level atom (without a recoil)
with a single-mode quantized field in an ideal cavity is known as the
Jaynes-Cummings model \cite{JC}. It describes the atom-field system as
a quantum-electrodynamical object whose evolution in time is
(quasi)periodic. In Schr\"odinger picture, it may be described by an
infinite set of linear ordinary differential equations for the probability
amplitudes to find the atom in the ground/excited state and the field in the
state with $n$ photons, where $n$ runs from zero to infinity. When adopting
the semiclassical approximation, we reduce, by hook or by crook, this
infinite set to a small number of equations for atomic and field variables.
In fact, we decouple the quantum atom-field system into an atom and field
parts which may exchange excitations with each other. In doing so, we get
automatically products of the atom and field variables in the equations of
motion. The respective Maxwell--Bloch equations may, under appropriate
conditions, produce  chaos in the classical sense of sensitive dependence on initial
conditions in the reduced classical phase space spanned by the atomic and
field expectation values. It resembles the procedure of deriving the famous
Lorenz equations from an infinite hierarchy of mode equations in fluid
convection by reducing it to three main modes only. 

The reduction of an
infinite set of linear equations, that are not chaotic in the classical sense, 
to a finite set of nonlinear equations, that may be chaotic, makes an impression
that it is only a mathematical trick. It seems that it is not only a useful
trick enabling us to handle with the equations of motion but the reduction is a
model of processes that may occur in nature. 
Coherence loss caused by inevitable interaction with
environment (decoherence) breaks quantum unitarity suppressing some quantum
properties of motion and revealing its classical properties. Such a situation
can be modeled with the use of a quantum-classical hybrid,  a system with
quantum and classical degrees of freedom dynamically coupled to each other. Which
part of the whole system under consideration is quantized and which one is
treated classically depends on the physical situation. If we have a large
number of atoms (or photons) we may adopt with a good accuracy the
semiclassical approximation treating the cavity field as a classical one.
When considering a single atom with a recoil in a high-quality microcavity it
is necessary to quantize the internal electronic atomic and field degrees of freedom but
the translational atomic motion may be treated classically if the values of
the atomic momentum are greater than the photon momentum.

\section {Atom without recoil interacting with a quantized radiation field in an ideal 
cavity}

The interaction between matter and radiation, commonly evidenced by 
spontaneous emission and induced emission and absorption of photons 
by atoms, is one of the most fundamental dynamical interactions in nature. 
In free space, an excited  atomic state decays irreversibly because an 
infinity of vacuum electromagnetic states is available  to the radiated 
photons which disappears in the free-space mode continuum.
The situation may be changes if the mode structure and the density of modes 
are modified by placing the radiating atom into a high-quality cavity. A
 cavity-induced enhancement and inhibition of spontaneous emission has been 
observed in a number of experiments in different ranges of the electromagnetic 
field, from microwaves to visible light (for a review see \cite{W92,RH95}). 

In a near-resonant ideal cavity with a single-mode frequency closed to the 
atomic transitional frequency, the radiation emitted by the atom is reflected 
at the cavity walls and is reabsorbed by the atom many times before it 
dissipates. In particularly, spontaneous emission becomes reversible. 
In cavity quantum electrodynamics near-resonant interaction of a single 
atom with a single-mode cavity field is commonly modeled by the Hamiltonian
\begin{equation}
\hat H=\hat H_a+\hat H_f+\hat H_{int},
\label{1}
\end{equation}
where
\begin{equation}
\hat H_a=\frac12\hbar\omega_a \hat \sigma_z
\label{2}
\end{equation}
represents the free motionless two-level atom with $\hbar\omega_a$ being 
the energy separation between the electronic states, the ground 
$\left|1\right>$ and 
excited $\left|2\right>$ ones. The Pauli spin operator acts on the states as
\begin{equation}
\hat \sigma_z\left|1\right>=-\left|1\right>,\quad \hat \sigma_z\left|2\right>=\left|2\right>.
\label{3}
\end{equation}
The Hamiltonian
\begin{equation}
\hat H_f=\hbar\omega_f(\hat a^\dag \hat a+\frac12)
\label{4}
\end{equation}
represents the mode of the isolated field. The photon annihilation $\hat a$ and 
creation $\hat a^\dag$ operators with the commutation rule $[\hat a, \hat a^\dag] = 1$ act on 
the photon number states as follows:
\begin{equation}
\hat a\left|n\right>=\sqrt{n}\left|n-1\right>,\quad \hat a^\dag\left|n\right>=\sqrt{n+1}\left|n+1\right>.
\label{5}
\end{equation}
The electric dipole interaction between the atom and the field mode is 
described by the operator
\begin{multline}
\hat H_{int}=\hbar\Omega_0(x)(\hat a+\hat a^\dag)(\hat \sigma_++\hat \sigma_-)
\simeq \\
\hbar\Omega_0(x)(\hat a\hat \sigma_++\hat a^\dag\hat \sigma_-),
\label{6}
\end{multline}
where $\Omega_0(x)$ is the atom-field coupling constant at the atomic 
position $x$ in the one-dimensional cavity, which is known under the names 
``vacuum or single-photon Rabi frequency''. 
The Pauli spin operators $\hat \sigma_\pm$ with the commutation rules 
$[\hat \sigma_\pm,\hat \sigma_z]=\mp 2\hat \sigma_\pm$, $[\hat \sigma_+,\hat 
\sigma_-]=\hat \sigma_z$ 
describe transitions between the ground and excited electronic states 
\begin{equation}
\hat \sigma_+\left|1\right>=\left|2\right>,\quad \hat \sigma_-\left|2\right>
=\left|1\right>,\quad \hat \sigma_+\left|2\right>=\hat \sigma_-\left|1\right>=0.
\label{7}
\end{equation}

The first form of the Hamiltonian (\ref{6}) follows from the quantization of 
the classical electric 
dipole atom-field interaction by replacing the atomic dipole moment by the 
operator $\hat \sigma_x=\hat \sigma_++\hat \sigma_-$  and the classical electric field 
by the operator $\hat a+\hat a^\dag$.
The products $\hat a^\dag\hat \sigma_-$ and $\hat a\hat \sigma_+$ describe the transition of the 
atom from the ground 
(excited) to the excited (ground) state and simultaneous annihilating 
(creating) a photon in the field mode, respectively. They are usual 
energy-conserving processes. The product $\hat a^\dag\hat \sigma_+$ describes the 
transition 
of the atom from the ground to the excited state and simultaneous creating 
a photon in the mode. The reverse process of simultaneous atom and field 
de-excitation is described by the term $\hat a\hat \sigma_-$. The last two processes 
are not energy-conserving and may be neglected while we deal with 
a single atom. Omitting the energy-nonconserving terms defines the 
Jaynes-Cummings model \cite{JC} which was proposed originally in 1963 as a 
purely theoretical tool for studying fundamentals of the interaction between 
matter and radiation. It seemed to be far from reality because neither the 
atom nor the cavity mode can hardly be perfectly isolated from environment 
with those-days technique. Exciting progress in experimental techniques has 
drastically changed the situation. Experiments, using very high-quality 
microwave cavities with $Q\sim 10^{10}$ and optical microcavities with 
$Q\sim10^6$, have now achieved 
the exceptional circumstance of strong coupling between atoms and a 
cavity field with the strength of the coupling exceeding the atomic and 
cavity decays that provides manipulations with single atoms and photons and 
experimental proving of some predictions of the Jaynes-Cummings model 
\cite{QED,W92,RBN,VWW}. 
The art of experimentalists and theoretical efforts have opened exciting 
perspectives for realizing gedanken experiments on fundamentals of quantum 
mechanics and implementing quantum communications and computing. 
As we will show in this lecture, an atom, interacting with a radiation field 
in a high-$Q$ cavity, provides an example of strongly-coupled microscopic 
nonlinear system whose dynamics may be very complicated and even chaotic. 

The Jaynes-Cummings Hamiltonian has the form  
\begin{multline}
\hat H_{JC}=\frac12\hbar\omega_a\hat \sigma_z+\hbar\omega_f(\hat a^\dag \hat a
+\frac12)+\\
\hbar\Omega_0f(x)(\hat a^\dag\hat \sigma_-+\hat a\hat \sigma_+),
\label{8}
\end{multline}
where $\Omega_0$ is an amplitude 
value of the vacuum Rabi frequency and $f(x)$ is a shape 
function of the cavity mode. Let us expand the atom-field  state vector 
over the electronic atomic states and Fock (or photon-number) field states 
$\left|n\right>$ 
\begin{equation}
\left|\Psi(t)\right>=\sum_{n=0}^\infty a_n(t)\left|2,n\right>+b_n(t)\left|1,n\right>,
\label{9}
\end{equation}
where $a_n(b_n)$ are probability amplitudes to find the atom in the 
excited (ground) state and $n$ photons in the field mode, respectively. 
Substitution of Eqs.(\ref{8}) and (\ref{9}) into the Schr\"odinger equation
\begin{equation}
i\hbar\frac{d\left|\Psi(t)\right>}{dt}=\hat H_{JC}\left|\Psi(t)\right>
\label{10}
\end{equation}
results in an infinite-dimensional set of coupled linear equations for the 
complex-valued probability amplitudes
\begin{equation}
\begin{aligned}
\dot a_n&=-i\left[\Delta_a a_n + \sqrt{n+1}\,f(x) b_{n+1}\right],&\\
\dot b_{n+1}&=\phantom{-}
i\left[\Delta_b b_{n+1}^*  + \sqrt{n+1}\,f(x) a_n^* \right],&\\
\end{aligned}
\label{11}
\end{equation}
where dot denotes differentiation with respect to dimensionless time 
$\tau=\Omega_0t$, $\Delta_a=(n\omega_f+\omega_a/2)/\Omega_0$, and 
$\Delta_b=[(n+1)\omega_f-\omega_a/2]/\Omega_0$. It is convenient to define new 
real-valued combinations of the probability amplitudes
\begin{equation}
\begin{aligned}
u_n=2{\rm Re}{\left(a_nb_{n+1}^* \right)},\quad
v_n=-2{\rm Im}{\left(a_nb_{n+1}^*\right)},\\
z_n=|a_n|^2-|b_{n+1}|^2,
\end{aligned}
\label{12}
\end{equation}
for which we get from (\ref{11}) the quantum Bloch-like equations of motion
\begin{equation}
\begin{aligned}
\dot u_n&=\delta v_n,\\
\dot v_n&=-\delta u_n-2\sqrt{n+1}\,f(x) z_n,\\
\dot z_n&=2\sqrt{n+1}\,f(x) v_n,\qquad n=0,\,1,\,2,\,\dots,
\end{aligned}
\label{13}
\end{equation}
where $\delta=(\omega_a-\omega_f)/\Omega_0$ is the dimensionless detuning 
between the atomic transition, $\omega_a$, and the field-mode, $\omega_f$, 
frequencies. For each specified photon number $n$, the quantity 
$R_n^2=u_n^2+v_n^2+z_n^2$ is conserved. Two global integrals of motion 
\begin{multline}
W_{JC}= \sum_{n=0}^\infty\sqrt{n+1}\,f(x) u_n -
\frac{\delta}{2}\sum_{n=0}^\infty z_n,\\ \sum_{n=0}^\infty R_n=1
\label{14}
\end{multline}
reflect conservation of the total energy and of the total probability, 
respectively. Eqs.(\ref{13}) can be easily solved for each $n$ in terms of 
trigonometric functions (we assume that $f(x)$ does not depend on time). 
The general solution of the Schr\"odinger equation is the 
sum of the solutions of an infinite set of the independent Bloch-like equations.
The exact general solution for one of the measured quantities, the atomic 
population inversion
\begin{equation}
z(\tau)=\sum_{n=0}^\infty z_n(\tau),
\label{15}
\end{equation}
is the following:
%
\begin{multline}
z(\tau)=\sum_{n=0}^\infty u_n(0)\frac{2\delta\sqrt{n+1}\,f(x)}{\Omega_n^2}
(1-\cos\Omega_n\tau)+\\
v_n(0)\frac{2\sqrt{n+1}\,f(x)}{\Omega_n}
\sin\Omega_n\tau +\\
z_n(0)\frac{\delta^2+4(n+1)\,f^2(x)\cos\Omega_n\tau}{\Omega_n^2},
\label{16}
\end{multline}
%
where 
\begin{equation}
\Omega_n=\sqrt{\delta^2+4(n+1)\,f^2(x)}
\label{17}
\end{equation}
is the $n$-photon Raby frequency, $u_n(0)$, $v_n(0)$ and $z_n(0)$ are the 
respective initial values 
which are determined by the initial atomic and field states. As it is seen from 
(\ref{16}), the solution with initially sharply defined atomic and field energy 
states ($\left|1\right>$ or $\left|2\right>$ and $\left|n\right>$) describes a periodic exchange of one 
quantum of 
energy between the atom and the field. 

The Fock state $\left|n\right>$ with a specified 
number of photons in the mode is an exotic field state. In general, a  pure 
quantized field state is an infinite superposition of the photon-number states
\begin{equation}
\left|f\right>=\sum_{n=0}^\infty c_n \left|n\right>,
\label{18}
\end{equation}
where $p_n=|c_n|^2$ is the probability for observing $n$ photons. The most 
classical of single-mode quantum states is a coherent state
\begin{equation}
\left|\alpha\right>=e^{-|\alpha|^2/2}\sum_{n=0}^\infty \frac{\alpha^n}{n!}\left|n\right>\equiv
\sum_{n=0}^\infty c_n(\alpha)\left|n\right>,
\label{19}
\end{equation}
whose photon probabilities follow a Poisson distribution with the mean number 
of photons $<n>=|\alpha|^2$  and the root mean square spread 
$\Delta n=|\alpha|$. Evolution of the 
population inversion of an atom, $z$, in the field, initially prepared to be 
in a coherent state, contains all the Rabi frequencies $\Omega_n$. The 
incommensurability of Rabi frequencies for different $n$ inevitably washes out 
the periodicity of population transfer resulting in a collapse of the 
excited-state occupation probability. Since the frequencies $\Omega_n$ form a 
\begin{figure}[ht]
\begin{center}
{\bf (a)} \parbox[ht]{0.45\textwidth}{\rule{0pt}{0pt}\\
\includegraphics[width=0.45\textwidth]{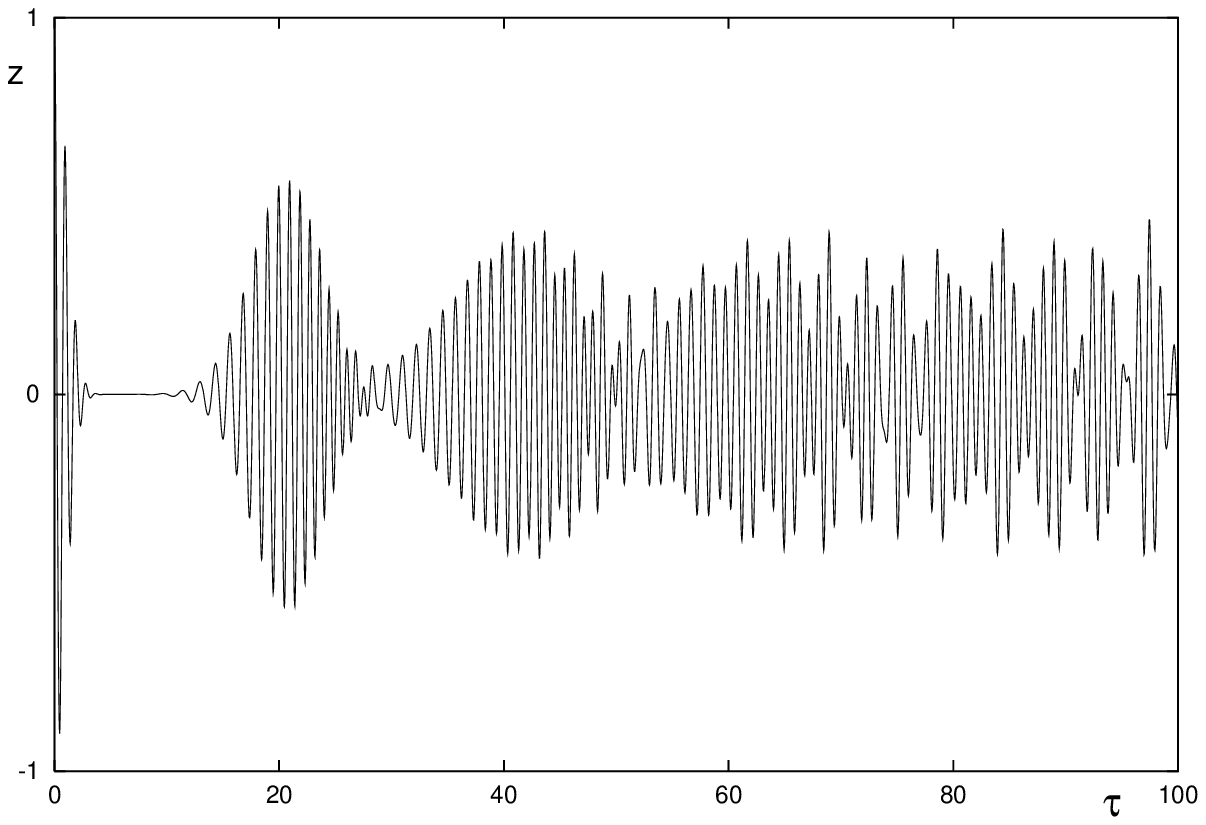}}\\
{\bf (b)} \parbox[ht]{0.45\textwidth}{\rule{0pt}{0pt}\\
\includegraphics[width=0.45\textwidth]{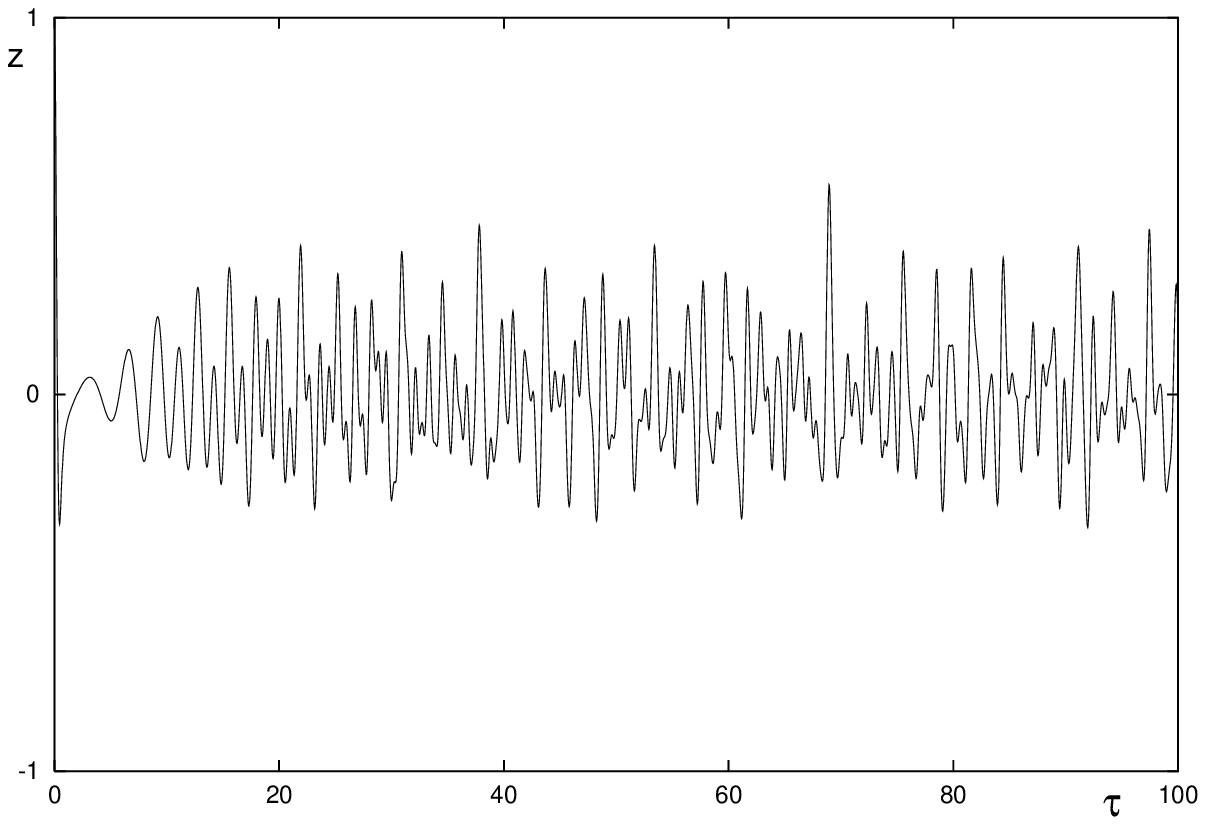}}\\
\end{center}
\caption
{Atomic population inversion $z(\tau)$ for the resonant Jaynes-Cummings 
model  with an initially excited motionless atom and (a)  coherent and 
(b) Bose-Einstein   
initial states of the cavity quantized field.}
\label{fig1}
\end{figure}
discrete set, the signal reappears after a time (see Fig. ~\ref{fig1}a computed 
with $f(x)=1$, $\delta =0$ and with the initial  conditions (\ref{33})). 
The collapse-revival phenomenon, a direct demonstration of quantum nature 
of the radiation field, has been demonstrated experimentally \cite {RWK,W92}. 

Let us consider as well another pure field state 
\begin{equation}
\left|\chi\right>=\sum_{n=0}^\infty c_n(\chi)\left|n\right>,
\label{20}
\end{equation}
whose photon probabilities follow a Bose-Einstein  distribution
\begin{equation}
p_n(\chi)=\frac{<n>^n}{(1+<n>)^{n+1}}. 
\label{21}
\end{equation}
Collapses and revivals also occur with this  field state 
(see Fig. ~\ref{fig1}b computed 
with $f(x)=1$, $\delta =0$ and with the initial  conditions (\ref{34})), 
but the collapse time 
is much shorter as compared with a coherent state because the Bose-Einstein 
spread in the photon number is far larger than for a coherent field with the 
same mean number of photons. 

Fig. ~\ref{fig1} demonstrates that the simple Jaynes-Cummings model with 
$f(x)=1$ may produce  
very complicated time evolution. However, the respective oscillations, of 
course, are not chaotic in the classical sense of exponential sensitivity to 
small changes in initial conditions. It is clear from the very structure of 
the equations of motion (\ref{13}) which are infinite-dimensional but linear. 
In classical mechanics the phase space is continuous and states of a 
classical system can be arbitrarily close to each other providing a possibility 
of exponential diverging of initially close trajectories in a bounded region 
of the phase space. In  quantum mechanics there is no notion of a 
trajectory, and the quantum phase space is not continuous due to the 
Heisenberg uncertainty principle. Evolution of {\it an isolated quantum system} 
is unitary and no dynamical chaos in the classical sense is possible with such 
systems.
The notion "quantum chaos" just refers to the behavior of quantum systems whose 
classical counterparts behave chaotically \cite{Chir79,Z81}. Real quantum systems 
are not isolated, they interact with their environment and, under attempts  
to measure their states, with  classical devices which, by virtue of their purpose, 
should be 
in unstable states. Loosing of coherence, due to inevitable interaction with
environment (decoherence), breaks down quantum unitarity suppressing  quantum 
properties of motion and manifesting classical ones. A quantum rotor with 
periodic kicks provides an illustrative example of suppressing quantum 
dynamical localization when one takes into account a thermal "bath" in 
the respective equations of motion \cite{Haake}. In the next section we will 
elaborate these ideas with an extended version of the strongly coupled 
atom-field system considered above.
        
\section{A quantum-classical atom-field hybrid}

In the process of emitting and absorbing photons of the 
cavity-field mode, the atom not only changes its internal electronic states but 
its external translational state is changed as well due to the photon recoil 
effect. In this section we consider a single two-level atom with mass $m_a$ 
moving 
in an ideal cavity which sustains a single standing-wave mode along the axis 
$x$ with the wave vector $k_f$ and the shape function 
$f(x)=-\cos(k_f\hat x)$. The respective 
Hamiltonian is the following extension of the Jaynes-Cummings Hamiltonian (
\ref{8})
\begin{multline}
\hat H=\frac{1}{2m_a}\hat p^2+\frac12\hbar\omega_a\hat\sigma_z+
\hbar\omega_f\hat a^\dag\hat a-\\
\hbar\Omega_0\left(\hat a^\dag\hat\sigma_-+
\hat a\hat\sigma_+\right)\,\cos{k_f\hat x},
\label{22}
\end{multline}
where the momentum $\hat p$ and position $\hat x$ operators satisfy the 
standard commutation relation $[\hat x, \hat p]=i\hbar$. 
Operators, belonging to different degrees of freedom, commute with 
each other at the same time moment. 

We have now three degrees of freedom, the internal atomic and the 
field ones, and the external (or translational) atomic degree of freedom. 
The first two degrees of freedom are treated as fully quantum ones in the 
Schr\"odinger picture. In fact, there are an infinite number of quantum degrees 
of freedom (see Eq. (\ref{13})) which are entangled. The external degree
of freedom will be treated as the classical one that may be justified by large 
values of the atomic momentum as compared with the photon momentum $\hbar k_f$. 
The Hamilton
equations of motion for the classical external degree of freedom is easily 
found from the 
Hamiltonian (\ref{22})
\begin{equation}
\frac{d<\hat x>}{dt}=\frac{\partial <\hat H>}{\partial <\hat p>},\quad
\frac{d<\hat p>}{dt}=-\frac{\partial <\hat H>}{\partial <\hat x>},
\label{23}
\end{equation}
where $<\dots>$ denotes an expectation value of the corresponding operator 
over a quantum state $\left|\Psi\right>$ of the electronic-field Hamiltonian. Using the 
normalizations $x=k_f<\hat x>$, $p=<\hat p>/\hbar k_f$, and $\tau =\Omega_0t$, 
we get from (\ref{23})
\begin{equation}
\begin{aligned}
\dot x=&{\phantom-}\alpha p,\\
\dot p=&-\left<\Psi(\tau)\left|\hat a^\dag\hat \sigma_-+\hat a\hat \sigma_+\right|\Psi(\tau)\right>\sin x,
\end{aligned}
\label{24}
\end{equation}
where $\alpha=\hbar k_f^2/m_a\Omega_0$ is the normalized recoil frequency 
which characterizes the average change in kinetic energy of the atom, 
$\hbar^2 k_f^2/2m_a$, in the process of emission and absorption of a photon. 
After computing the expectation value in (\ref{24}) with the state 
vector (\ref{9}), we obtain our basic Hamilton-Schr\"odinger equations
\begin{equation}
\begin{aligned}
\dot x_{\phantom n}&=\alpha p,\\
\dot p_{\phantom n}&=-\sum_{n=0}^\infty\sqrt{n+1}\, u_n\sin x,\\
\dot u_n&=\delta v_n,\\
\dot v_n&=-\delta u_n+2\sqrt{n+1}\, z_n\cos x,\\
\dot z_n&=-2 \sqrt{n+1}\, v_n\cos x, \qquad n=0,\,1,\,2,\,\dots
\end{aligned}
\label{25}
\end{equation}
which describe the atom-field quantum-classical hybrid with the quantum 
degrees of freedom dynamically coupled to the classical degree of freedom. 
This infinite set of {\it nonlinear ordinary differential equations} 
possesses an infinite number of the integrals of motion, the total energy 
integral
\begin{equation}
W=\frac{\alpha}{2}\,p^2-\sum_{n=0}^\infty\sqrt{n+1}\,u_n\cos x -
\frac{\delta}{2}\sum_{n=0}^\infty z_n,
\label{26}
\end{equation}
the Bloch-like integral for each $n$
\begin{equation}
R_n^2=u_n^2+v_n^2+z_n^2,
\label{27}
\end{equation}
and the global integral
\begin{equation}
\sum_{n=0}^\infty R_n=1.
\label{28}
\end{equation}
The infinite-dimensional nonlinear dynamical system (\ref{25}) is a
quantum generalization of the five-dimensional semiclassical set of the equations 
of motion for the same problem that was derived in Refs \cite {PK01,PS01} 
with a classical field.
The later one was shown \cite {PK01,PS01,P02,PEZ02} to be chaotic with positive
values of the maximal Lyapunov exponent in some ranges of the system's control
parameters. It is rather difficult to compute this quantitative indicator
of dynamical chaos with an infinite dimensional set of ODE's. However, we have
found another signatures of Hamiltonian chaos with Eqs. (\ref{25}), the main
of which are fractals which will be demonstrated in the next section.

\begin{figure}[ht]
\begin{center}
{\bf (a)} \parbox[ht]{0.45\textwidth}{\rule{0pt}{0pt}\\
\includegraphics[width=0.45\textwidth]{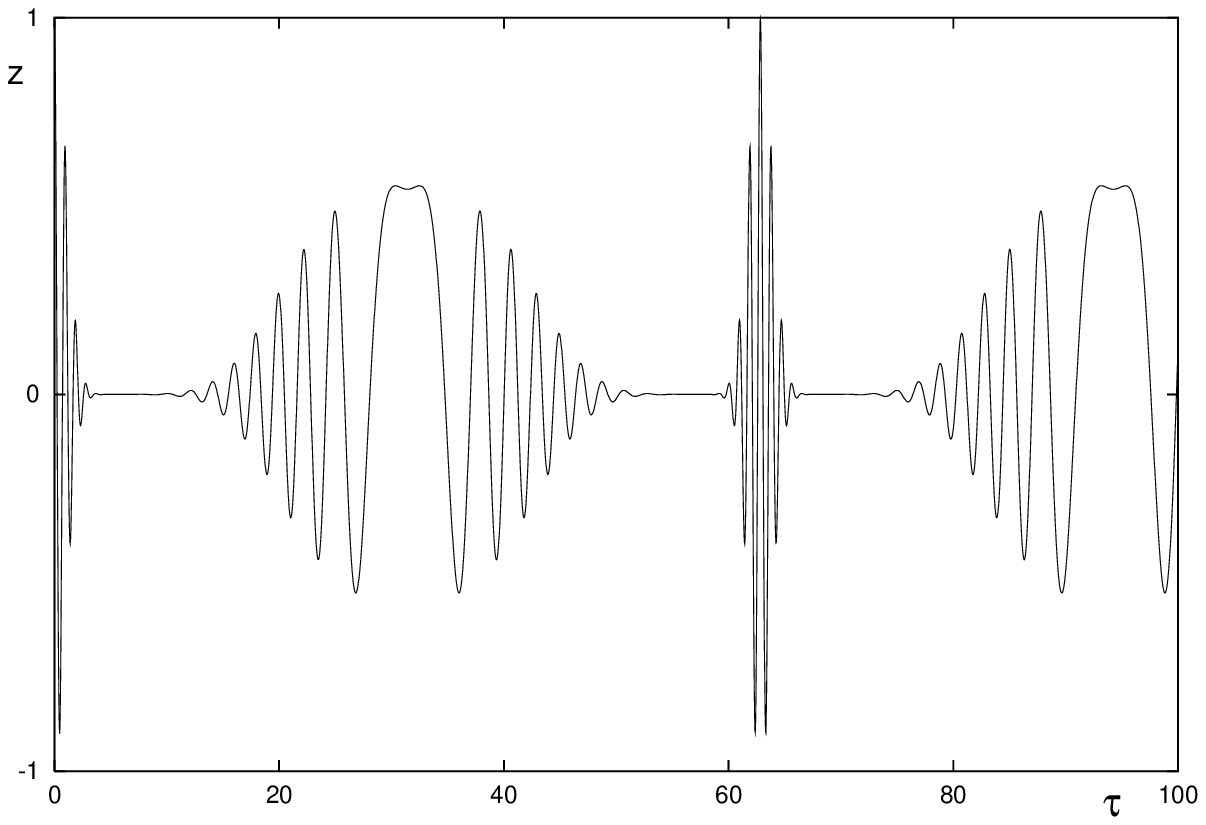}}\\
{\bf (b)} \parbox[ht]{0.45\textwidth}{\rule{0pt}{0pt}\\
\includegraphics[width=0.45\textwidth]{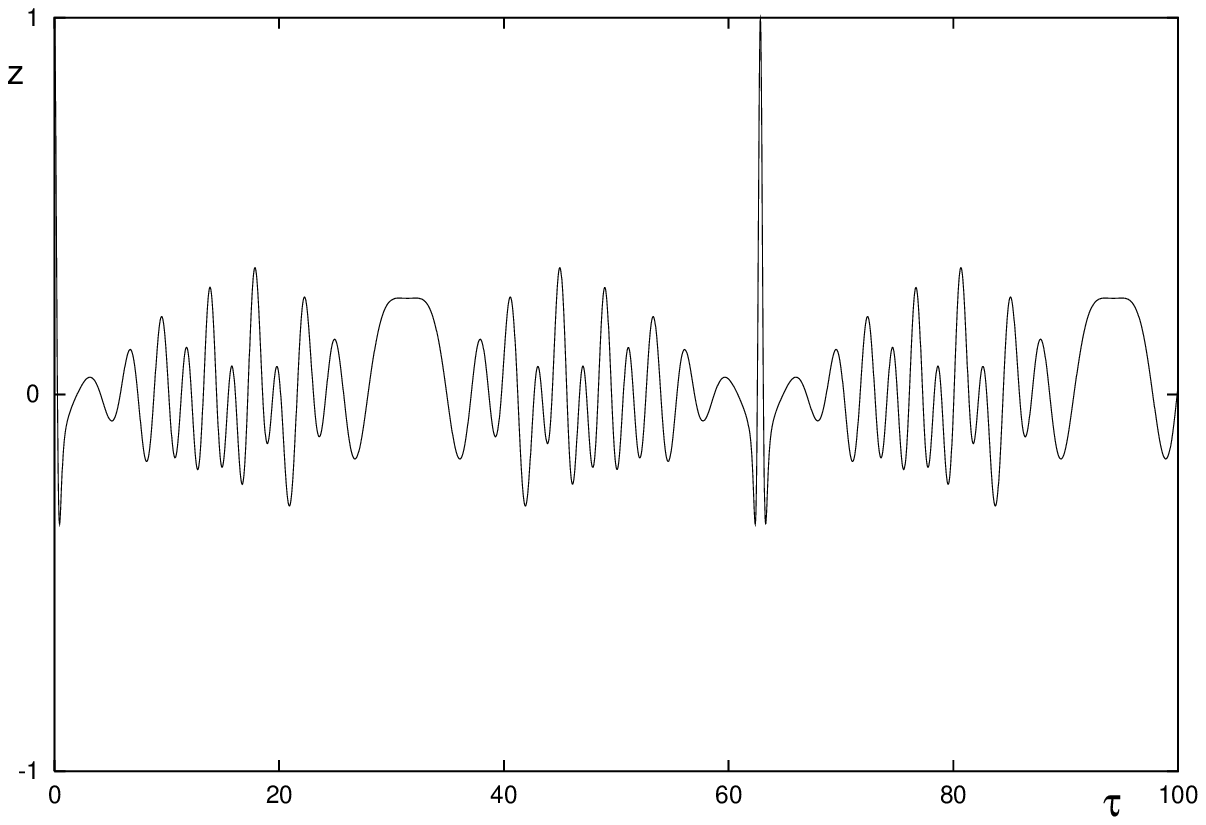}}\\
\end{center}
\caption
{The same as in FIG.~\ref{fig1} but with the resonant quantum-classical hybrid.}
\label{fig2}
\end{figure}
         
Let us compare the Rabi-oscillation signals with the atom-field quantum-classical 
hybrid (\ref{25}) and with the fully quantum Jaynes-Cummings model 
without recoil (\ref{13}). In Fig.~\ref{fig2}a and b the population 
inversion $z$ 
(see Eq. (\ref{15})) for an initially excited atom interacting in resonance with the
field that is initially in a coherent state (\ref{19}) and in a 
Bose-Einstein state
(\ref{20}), respectively, is shown. This figure should be compared with 
Fig.~\ref{fig1}
where the Rabi oscillations with the Jaynes-Cummings model have been computed 
for the same initial conditions and the field states with $<n> = 10$ but 
with a motionless atom. 
More pronounced collapses and revivals occur with the quantum-classical  
hybrid as compared with the fully quantum model without recoil. 
In order to understand peculiarities of the Rabi oscillations in Fig.~\ref{fig2} 
look 
at  Fig.~\ref{fig7}a demonstrating the scheme of gedanken experiments. An atom starts 
at $x=0$ and moves to the right with the initial momentum $p_0=50$. It reaches 
the first node of the standing wave, where the coupling coefficient with 
the field mode is zero at time moment $\tau_1=\pi/2\alpha p_0\simeq 31.4$ 
in dimensionless units (we 
choose $\alpha=0.001$ in our computer simulations). For both the field states, 
one can see the respective slowing down of the Rabi oscillations at the 
moments of time $\tau_s=(1+2s)\pi/2\alpha p_0$, when the atom transverses 
the $s$-th node. The 
pronounced peaks of revivals of the Rabi oscillations occur at the moments 
$\tau_r=r\pi/\alpha p_0$, 
when the atom transverses the $r$-th antinode of the standing wave where its 
coupling  with the field is maximal.
        
In spite of a complicated character of the Rabi oscillations, shown in
Fig.~\ref{fig2}, they are regular because at exact resonance, $\delta=0$, 
all the values
of the real-valued amplitudes $u_n$ are conserved during the evolution. 
With initially excited atom we have $u_n(0)=v_n(0)=0$ for each $n$. It 
immediately follows from
Eqs.(\ref{25}) that the atom moves with a constant velocity through the cavity.
It means that $x$ is a linear function of time, and the Hamilton-Schr\"odinger 
equations(\ref{25}) reduce to the periodically modulated linear Bloch-like
equations (\ref{13})  with  periodic solutions. The exact solution for the 
atomic population inversion can be easily found 
\begin{equation}
z(\tau) = \sum_{n=0}^\infty z_n(0)\cos \left(\frac{2\sqrt{n+1}}{\alpha p_0}\sin \alpha
p_0\tau \right). 
\label{28a}
\end{equation}
It is a periodic function (shown in Fig.~\ref{fig2}) with the period to be 
equal to $\pi/\alpha p_0$ and the maxima at $\tau_r =r\pi/\alpha p_0$ $(r=0,1,2, ...)$.

Out off resonance, $\delta\neq 0$, the quantum-classical hybrid may 
demonstrate chaos.
To diagnose chaos it is instructive to compute the maximal Lyapunov exponent 
$\lambda$ 
whose values depend on initial conditions, on the detuning $\delta$, the mean 
number of photons in the mode, and on the recoil frequency $\alpha$. If the 
quantized field is initially prepared in the Fock state $\left|n\right>$ with exactly $n$ 
quanta in the mode, 
the infinite-dimensional set (\ref{25}) reduces to 8 equations
\begin{equation}
\begin{aligned}
\dot x_{\phantom{n-1}}&=\alpha p,\\
\dot p_{\phantom{n-1}}&=-\left(\sqrt{n}\,u_{n-1}+\sqrt{n+1}\,u_n\right)\sin x,\\
\dot u_{n-1}&=\delta v_{n-1},\\
\dot v_{n-1}&=-\delta u_{n-1}+2\sqrt{n}\,z_{n-1}\cos x,\\
\dot z_{n-1}&=-2\sqrt{n}\,v_{n-1}\cos x,\\
\dot u_{n\phantom{-1}}&=\delta v_n,\\
\dot v_{n\phantom{-1}}&=-\delta u_n+2\sqrt{n+1}\,z_n\cos x,\\
\dot z_{n\phantom{-1}}&=-2\sqrt{n+1}\,v_n\cos x
\end{aligned}
\label{29}
\end{equation}
with initial conditions 
%
\begin{equation}
\begin{aligned}
&x(0)=x_0,\quad p(0)=p_0,\\
&z_{n-1}(0)=-\bigl|b_n(0)\bigr|^2,\quad z_n(0)=\bigl|a_n(0)\bigr|^2,\\
&u_{n-1}(0)=u_n(0)=v_{n-1}(0)=v_n(0)=0
\end{aligned}
\label{30}
\end{equation}
%
describing the atom-field system initially prepared in a superposition state
\begin{equation}
\begin{aligned}
&|\Psi(0)>=a_n(0)|2,\,n>+b_n(0)|1,\,n>,\\
&\bigl|a_n(0)\bigr|^2+\bigl|b_n(0)\bigr|^2=1.
\end{aligned}
\label{31}
\end{equation}
\begin{figure}[ht]
\begin{center}
\parbox[ht]{0.45\textwidth}{\rule{0pt}{0pt}\\
\includegraphics[width=0.45\textwidth]{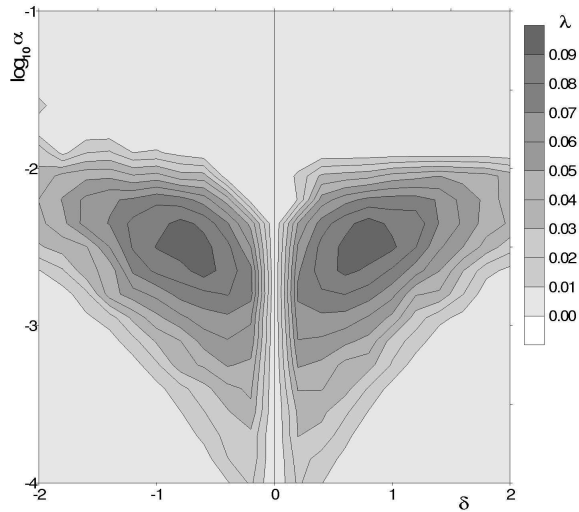}}
\end{center}
\caption
{The maximal Lyapunov exponent $\lambda$ with the quantum-classical hybrid 
in the Fock quantized field versus the atom-field detuning
$\delta$ and the logarithm of the dimensionless recoil frequency
$\alpha$.}
\label{fig3}
\end{figure}
The so-called topographic $\lambda$-maps, showing by color modulation values 
of $\lambda$ 
in the ranges of values of two control parameters with the third one to be
fixed, have been computed with the set (\ref{29}). In Fig.~\ref{fig3} we show 
the $\lambda$-map with initial zero atomic population, $a_n(0)=b_n(0)=1/
\sqrt 2$,  
$z(0)=z_{n-1}(0)+z_n(0)=0$, in the ranges of the detuning $\delta$ 
and the recoil frequency $\alpha$ with the fixed value of the initial 
number of photons $n=10$. At exact resonance, the set (\ref{29}) is 
integrable and $\lambda=0$ at $\delta=0$. In the range 
$\alpha\sim10^{-4}\div10^{-2}$, that corresponds to realistic values of the 
recoil frequency,
the maximal Lyapunov exponent may be positive and one may expect chaotic atomic
motion in the respective ranges of $\alpha$ and $\delta$.
\begin{figure}[ht]
\begin{center}
\parbox[ht]{0.45\textwidth}{\rule{0pt}{0pt}\\
\includegraphics[width=0.45\textwidth]{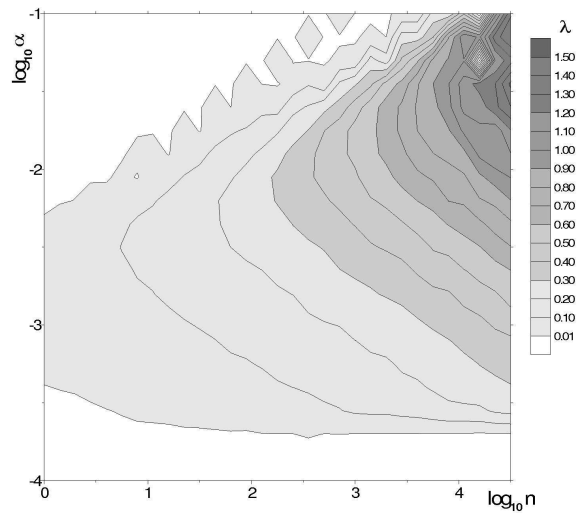}}
\end{center}
\caption{The double logarithmic plot of $\lambda$ versus $\alpha$ and initial number 
of photons in the Fock field $n$.}
\label{fig4}
\end{figure}
Another 
$\lambda$-map with initially excited atom, $a_n(0)=1, b_n(0)=0$, $z(0)=1$, 
demonstrates in 
Fig.~\ref{fig4} the values of $\lambda$ in  dependence on $\alpha$ and $n$ 
at $\delta=0.5$ in double logarithmic scale. The magnitude of the maximal 
Lyapunov exponent grows, in average, with increasing the initial number of
photons in the cavity mode.

\begin{figure}[ht]
\begin{center}
{\bf (a)} \parbox[ht]{0.45\textwidth}{\rule{0pt}{0pt}\\
\includegraphics[width=0.45\textwidth]{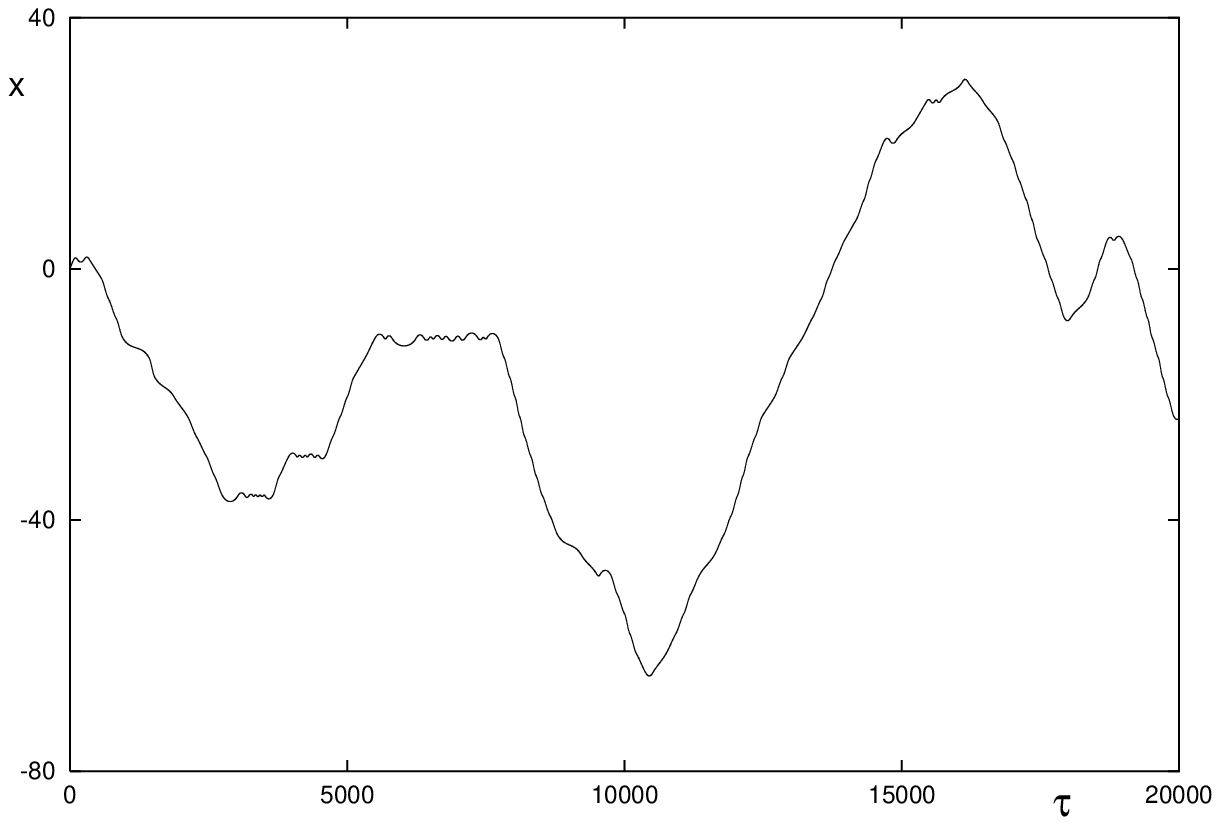}}\\
{\bf (b)} \parbox[ht]{0.45\textwidth}{\rule{0pt}{0pt}\\
\includegraphics[width=0.45\textwidth]{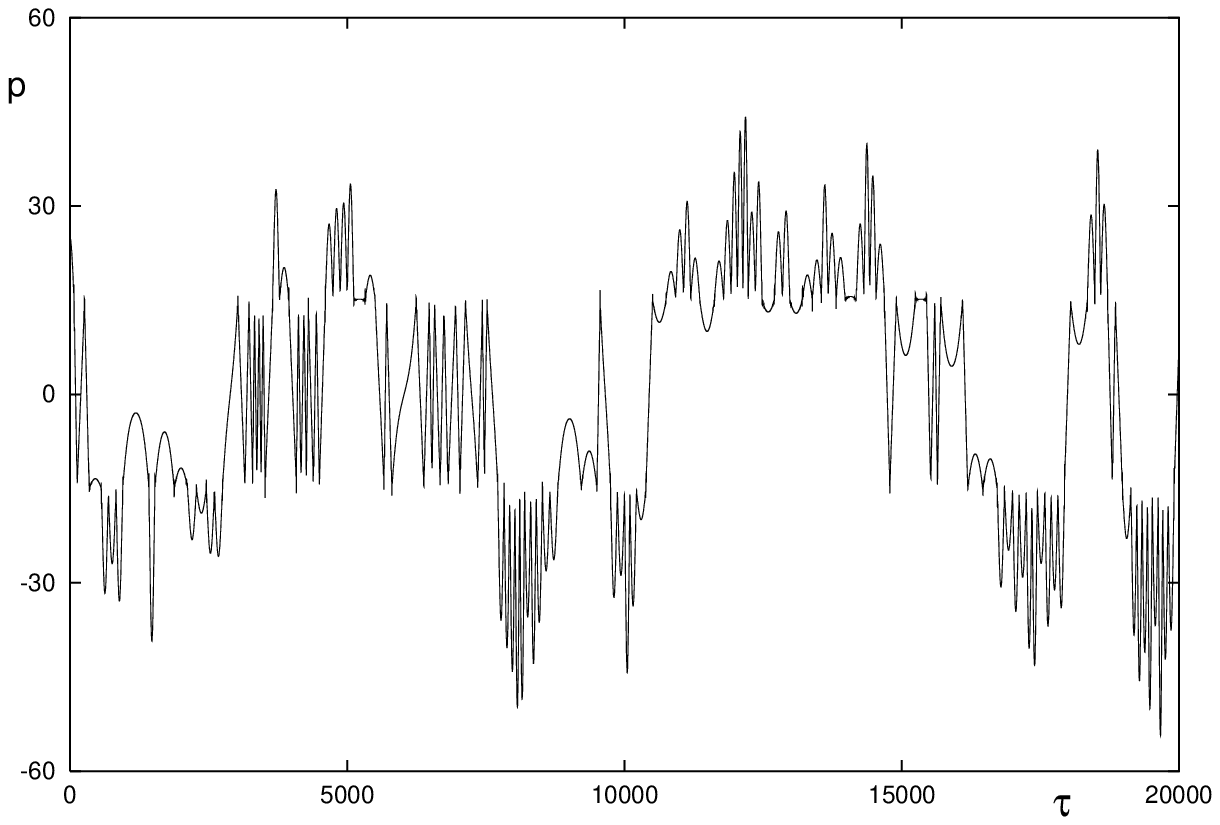}}
\end{center}
\caption{(a) A chaotic atomic trajectory with an initially coherent field and 
(b) the respective oscillations of the atomic momentum.}
\label{fig5}
\end{figure}
        
Computing the values of $\lambda$ is difficult with the cavity field initially 
prepared in a superposition state (\ref{18}) (including coherent and 
Bose-Einstein states) that generates an infinite number of the equations of motion 
(\ref{25}). We have found another signatures of chaos. 
In Fig.~\ref{fig5} an atomic trajectory, $x(\tau)$, and oscillations of the 
atomic momentum, $p(\tau)$,
are shown to illustrate manifestations of chaos in the set (\ref{25}) with 
the initially coherent field and the excited atom with $<n>=10$, $\delta=0.4$, 
$\alpha=0.001$, and $p_0=25$. A typical weakly chaotic atomic trajectory 
represents a kind of a random walking with small oscillations of the atom in 
potential wells, long ballistic flights of the L\'evy type with almost constant 
velocity and erratic turnbacks. Just the presence of L\'evy flights 
fluenses strongly the statistical properties of the system generating power-like 
statistical laws that have been extensively studied in the semiclassical 
approximation in Refs.\cite{PEZ02,AP}.
\begin{figure}[ht]
\begin{center}
{\bf (a)} \parbox[ht]{0.45\textwidth}{\rule{0pt}{0pt}\\
\includegraphics[width=0.45\textwidth]{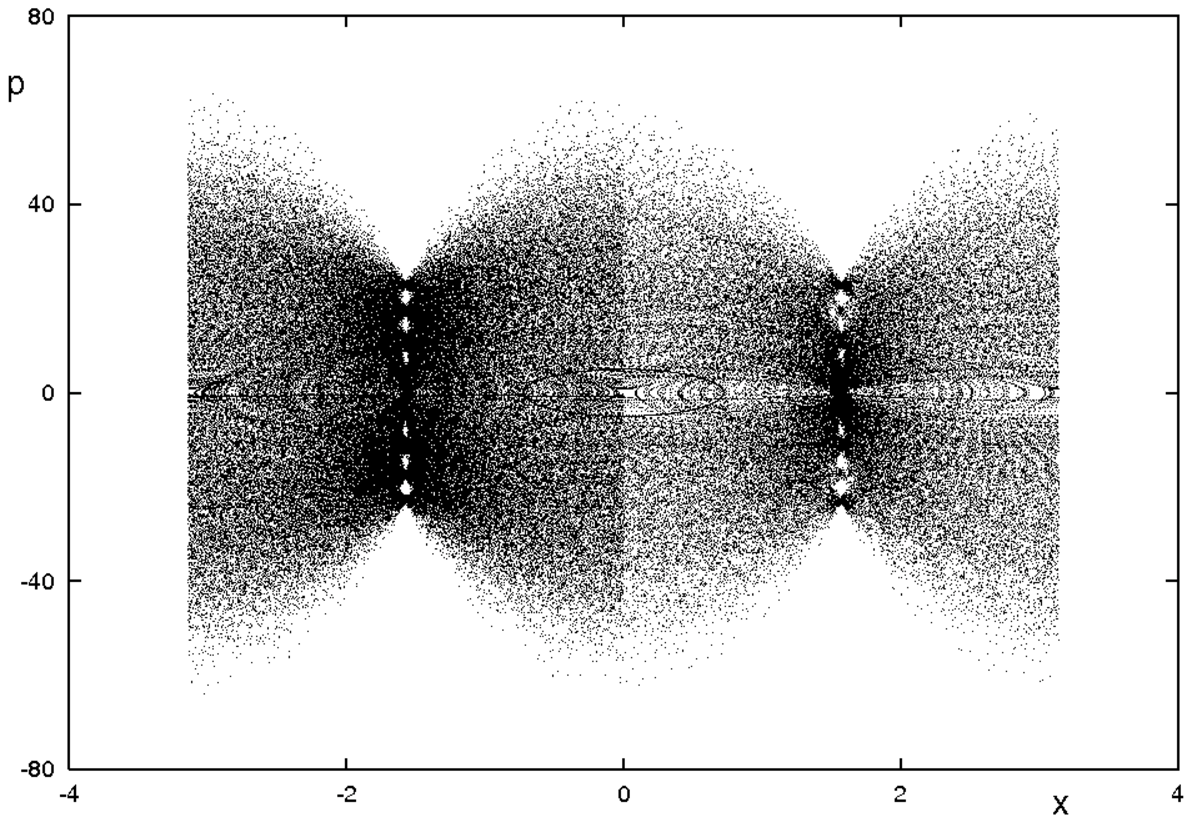}}\\
{\bf (b)} \parbox[ht]{0.45\textwidth}{\rule{0pt}{0pt}\\
\includegraphics[width=0.45\textwidth]{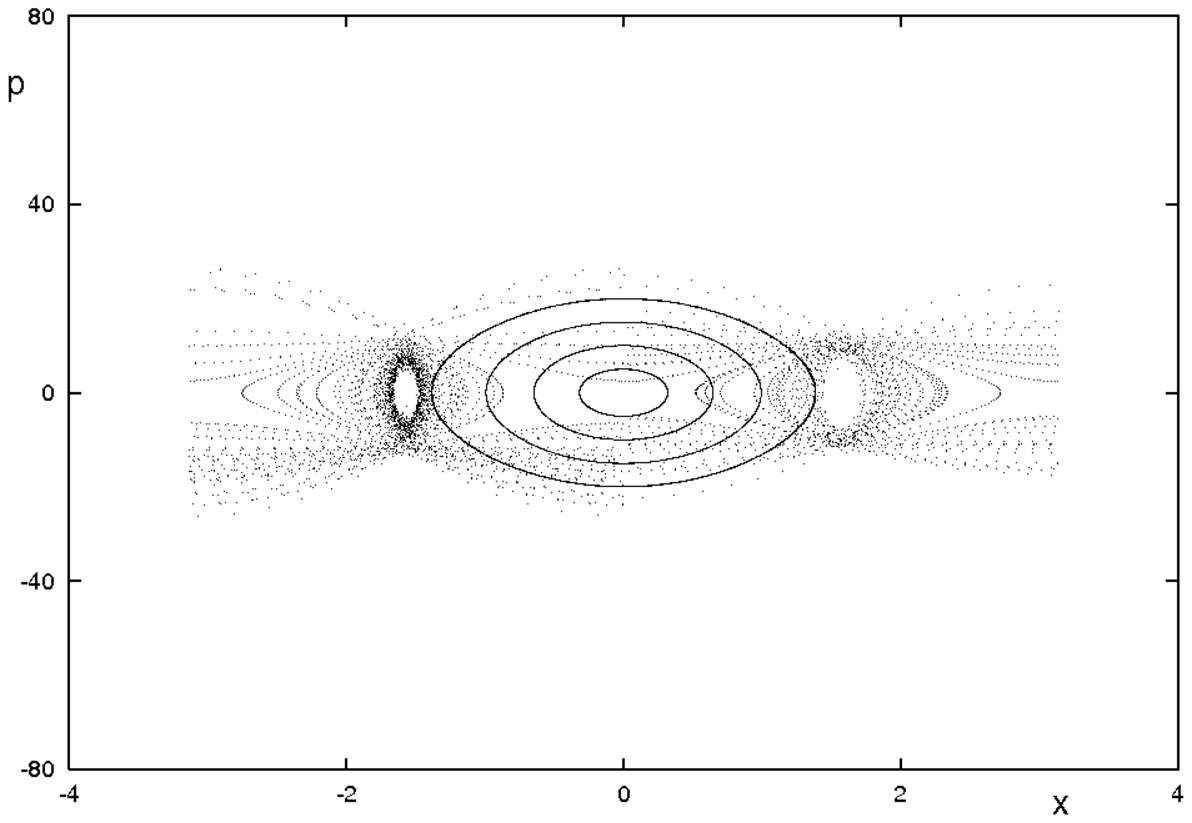}}
\end{center}
\caption
{Projections of the Poincar\'{e} sections on the plane of the atomic momentum
$p$ and the position $x$.
(a) Chaos  at $\delta = 0.1$ and (b) regular motion at $\delta = 0.5$.}
\label{fig6}
\end{figure}
Fig.~\ref{fig6}             
presents Poincar\'e sections of motion in the set (\ref{25}) with five different 
values of the initial atomic momentum to be 
projected on the plane of the atomic external variables $(x,p)$. 
Fig.~\ref{fig6}a demonstrates
an example of chaotic Poincar\'e section $(\delta=0.1)$ whereas Fig.~\ref{fig7}b 
presents a regular 
motion under the same conditions except for the detuning $\delta=0.5$. 

A feasible scheme for detecting manifestation of chaos with hot two-level            
Rydberg atoms moving in a high-Q microwave cavity has been proposed in               
\cite{P02}. The same idea could be realized with cold usual atoms in a high-Q        
microcavity. Consider a 2D-geometry of a gedanken experiment                         
with a monokinetic atomic beam propagating almost                                    
perpendicularly to the cavity axis $x$. In a reference frame moving with a           
constant velocity in the perpendicular direction, there remains only the transverse  
atomic motion along the axis $x$. One measures atomic population inversion           
after passing the interaction zone. Before injecting atoms in the cavity, it         
is necessary to prepare all the atoms in the same electronic state, say, in          
the excited state, with the help of a $\pi$-pulse of the laser radiation. It         
may be done with {\it only a finite accuracy}, say, equal to $\Delta                 
z_\text{in}$ for the initial population inversion $z_\text{in}$. The values          
of the population inversion $z_\text{out}$ are measured with detectors at a          
fixed time moment. If we would work with the values of the control parameters        
corresponding to the regular atom-field dynamics, we would expect to have a          
regular curve $z_\text{out}$--$z_\text{in}$. In the chaotic regime,  the             
atomic inversion at the output can be predicted (within a certain confidence         
interval $\Delta z$)  for a time not exceeding the so-called predictability          
horizon                                                                              
\begin{equation}                                                                     
\tau_p\simeq\frac{1}{\lambda}\ln\frac{\Delta z}{\Delta z_\text{in}},                 
\label{predtime}                                                                     
\end{equation}                                                                       
which depends weakly on $\Delta z_\text{in}$ and $\Delta z$. Since the               
maximal confidence interval lies in the range $|\Delta z|\leqslant 1$ and            
$\lambda$ may reach the values of the order of $1.5$ (see $\lambda$-maps),           
the predictability horizon in accordance with the formula (\ref{predtime})           
can be very short: with $\lambda$ =0.5 the predictability horizon 
$\tau_p$ may be of the order of 10 in units   
reciprocal of the vacuum Rabi frequency $\Omega_0$ that corresponds to                
$t_p\simeq 10^{-7}$ s with the realistic value of $\Omega_0\simeq 10^8$              
rad$\cdot$ s${}^{-1}$.                                                     
                                                                                     
\begin{figure}[ht]                                                                  
\begin{center}                                                                       
\includegraphics[width=0.45\textwidth]{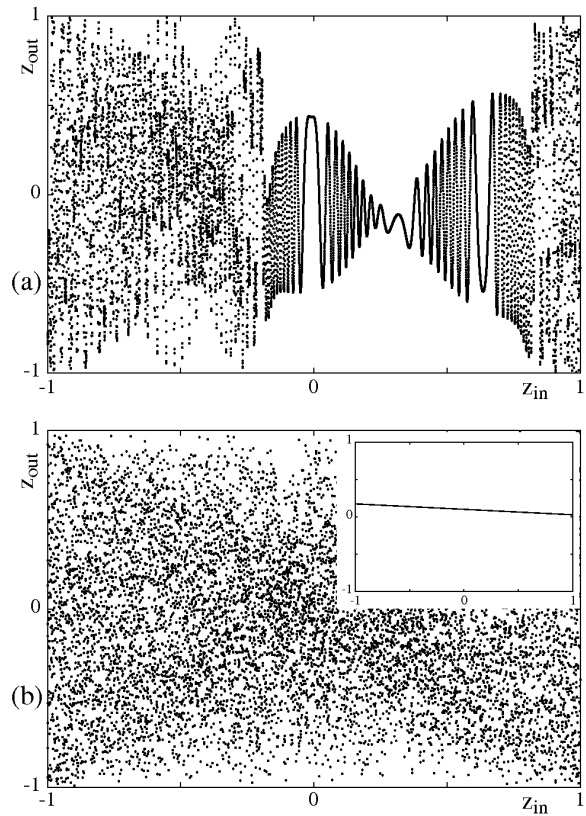}                                      
\caption{Dependence of the output values of the atomic population inversion          
$z_\text{out}$ on its initial values $z_\text{in}$ with $\delta=0.4$ (a)             
at $\tau=100$ and (b) at $\tau=200$ with the inset showing this dependence           
at the exact atom-field resonance, $\delta=0$.}\label{fig7}                          
\end{center}                                                                         
\end{figure}                                                                         
In the regular regime, the inevitable errors in preparing                            
$\Delta z_\text{in}$ produce the output errors $\Delta z_\text{out}$ of the          
same order. In the chaotic regime, the initial uncertainty increases                 
exponentially resulting in a complete uncertainty of the detected population         
inversion in a reasonable time. It is demonstrated in Fig.~\ref{fig7}, where         
we plot the dependence of the values of $z(\tau)=z_\text{out}$ at $\tau=100$         
(Fig.~\ref{fig7}a) and $\tau=200$ (Fig.~\ref{fig7}b) on the values of                
$z(0)=z_\text{in}$ in the chaotic regime  with an initially Fock field at            
 $\delta=0.4$ and  $\lambda\simeq0.05$. Simulation shows that an initial error $\Delta                            
z_\text{in}=10^{-4}$ in preparing the atomic electronic state leads to 
 complete uncertainty $\Delta       
z_\text{out}\simeq 2$ in a rather short time. To see                       
the difference, it is desirable to carry out a control experiment at the             
exact resonance ($\delta=0$) when the atomic motion is fully regular with any        
initial values. The dependence $z_\text{out}$--$z_\text{in}$ with $\delta=0$         
is demonstrated in the inset in Fig.~\ref{fig7}b with all the other control          
parameters and initial values being the same.

\section{Atomic dynamical fractals}

In this section, we treat the atom-photon interaction in a high-$Q$ cavity as a
chaotic scattering problem.
\begin{figure}[ht]
\begin{center}
{\bf (a)} \parbox[t]{0.45\textwidth}{\rule{0pt}{0pt}\\
\begin{center}
\includegraphics[width=0.35\textwidth]{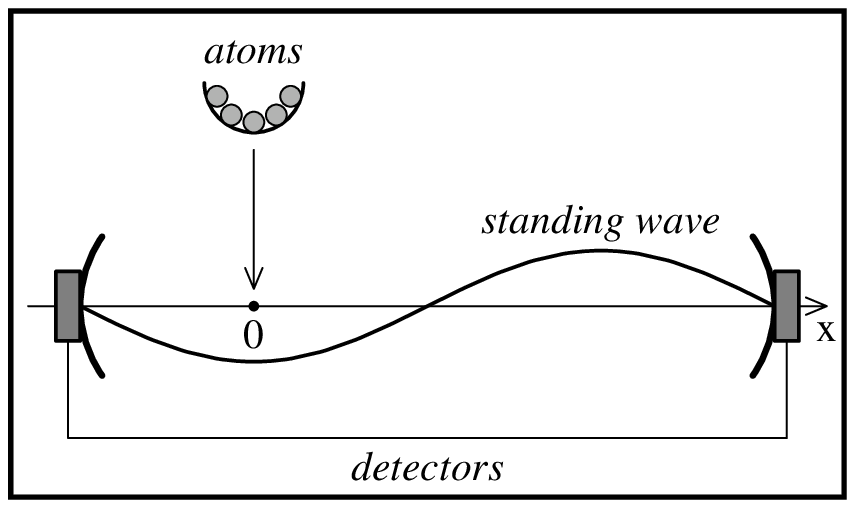}
\end{center}}\\
{\bf (b)} \parbox[t]{0.45\textwidth}{\rule{0pt}{0pt}\\
\includegraphics[width=0.45\textwidth]{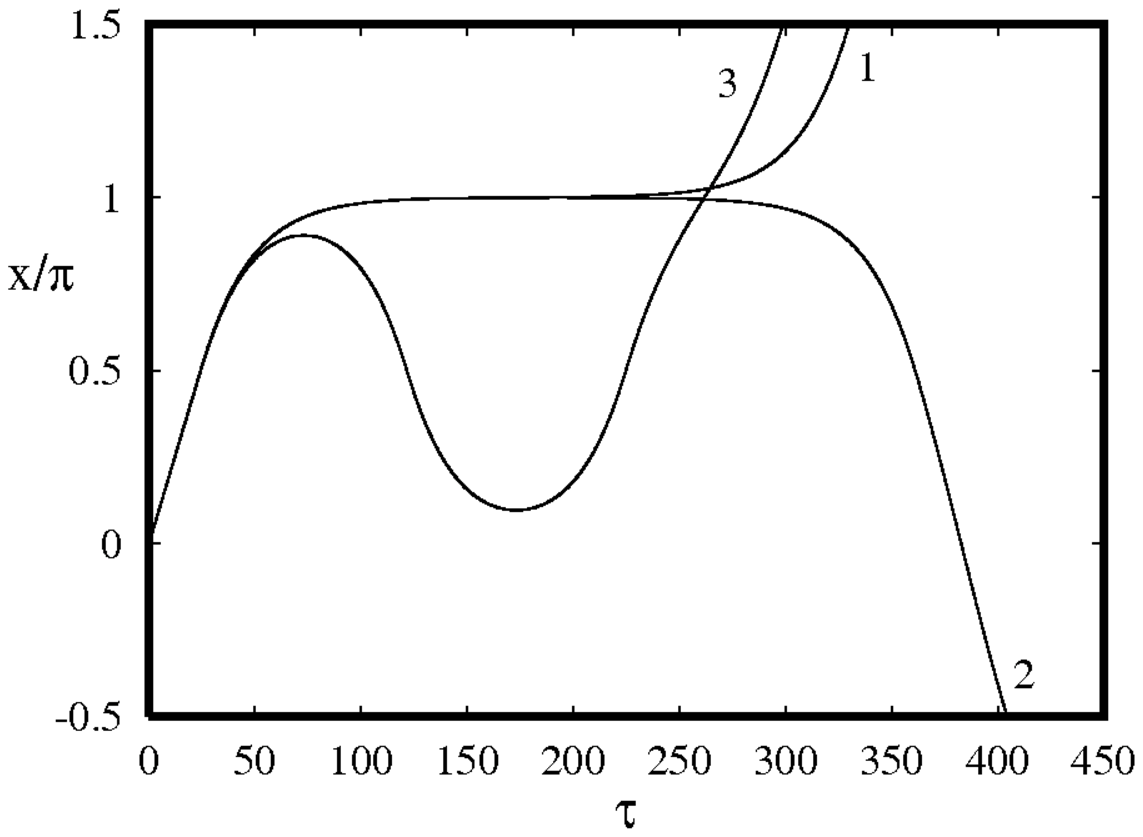}}
\end{center}
\caption{(a) Schematic diagram showing scattering of atoms at the standing
wave  and (b) sample atomic trajectories.}
\label{fig8}
\end{figure}
Let
us consider the scheme of scattering of atoms by the standing wave shown in
Fig.~\ref{fig8}a. Atoms, one by one, are placed at the point $x=0$ with
different initial values of the momentum $p_0$ along the cavity axis. For
simplicity, we suppose that they have no momentum in the other directions
(1D-geometry). We compute the time the atoms need to reach one of the
detectors placed at the cavity mirrors. The dependence of this exit time $T$
on the initial atomic momentum $p_0$ is studied under the other initial
conditions and parameters being the same.
To avoid complications that are not essential to the
main theme of this section, we consider the cavity with only two standing-wave
lengths.

\subsection {Fock fractal}

In this subsection the field is supposed to be initially prepared in a Fock state. 
Before injecting into a cavity, atoms are suppose to be prepared in
the superposition state with $u_{n-1}(0)= u_{n}(0)=v_{n-1}(0)=v_{n}(0)=0$,
$z_{n-1}(0)=-1/2, z_{n}(0)=1/2$, i.~e.\ in the state with zero population
inversion $z(0)=z_{n-1}(0)+z_{n}(0)=0$.

At exact resonance ($\delta=0$) with $u_{n-1}(\tau)=u_{n}(\tau)=0$,
the optical potential $U=(\sqrt{n}\,u_{n-1}+\sqrt{n+1}\,u_n)\cos x-
\delta(z_{n-1}+z_{n})/2$ is equal to zero, and
the analytical expression for the dependence in
question can be easily found to be the following:
\begin{equation}
\begin{aligned}
T(\delta=0)&=3\pi/2\alpha p_0 \text{ if } p_0>0,\\
T(\delta=0)&={\phantom{3}}\pi/2\alpha p_0 \text{ if } p_0<0.
\end{aligned}
\label{32}
\end{equation}
Atoms simply fly through the cavity in one direction with their initial constant
velocity and are registered by one of the detectors.
Out off resonance ($\delta\ne 0$), the atomic motion has been numerically found
in preceding section to be chaotic with
positive values of the maximal Lyapunov exponent in the following ranges of 
the values of the control parameters: the detuning  $\mid\delta\mid\ \lesssim 2$ 
and the recoil frequency $\alpha \simeq 10^{-4} \div 10^{-2}$. 
\begin{figure}[ht]
\begin{center}
{\bf (a)} \parbox[t]{0.45\textwidth}{\rule{0pt}{0pt}\\
\includegraphics[width=0.45\textwidth]{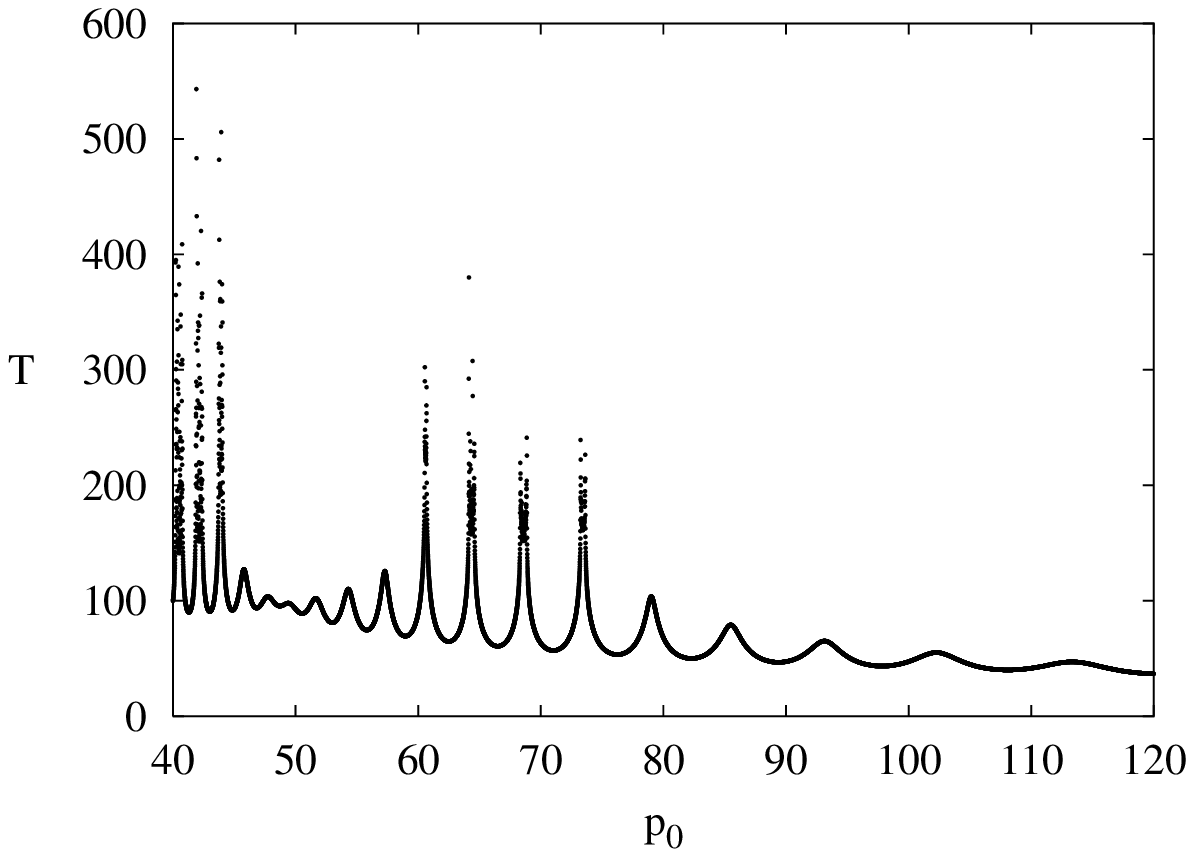}}\\
{\bf (b)} \parbox[t]{0.45\textwidth}{\rule{0pt}{0pt}\\
\includegraphics[width=0.45\textwidth]{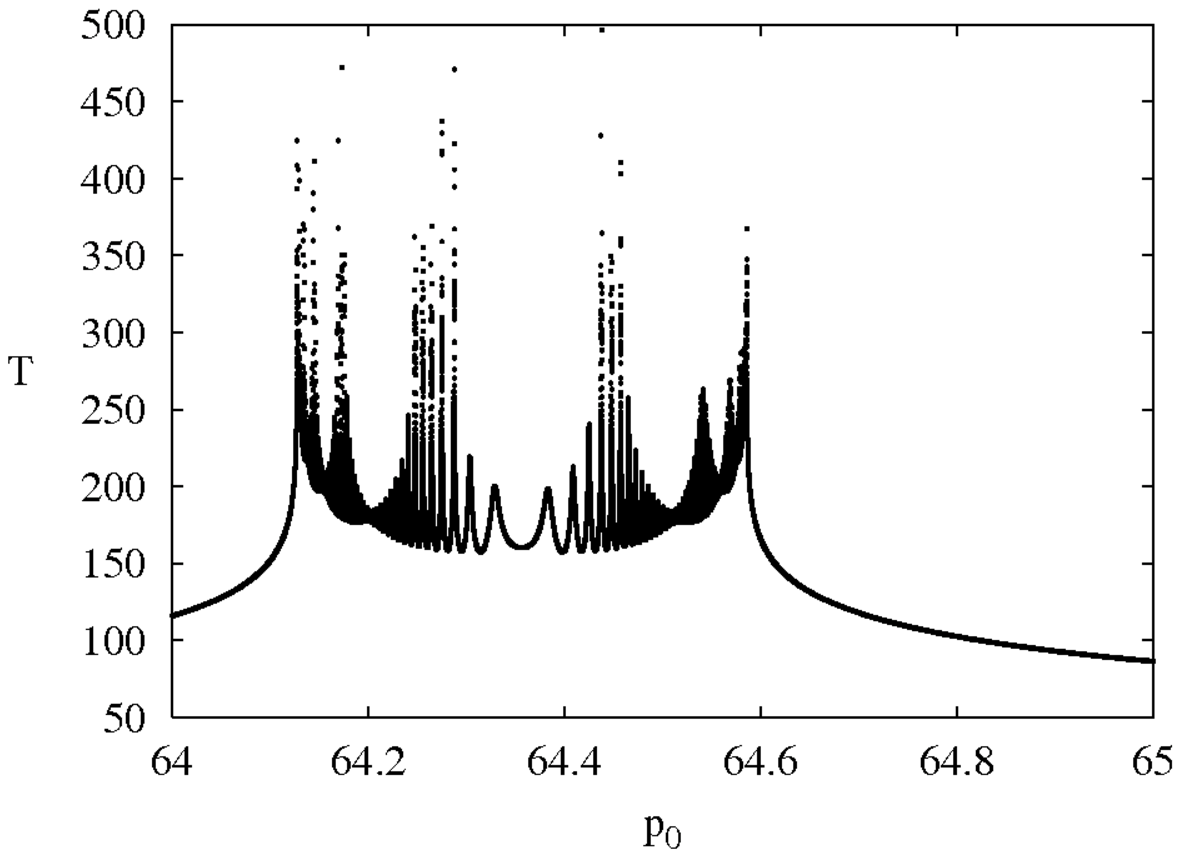}}\\
{\bf (c)} \parbox[t]{0.45\textwidth}{\rule{0pt}{0pt}\\
\includegraphics[width=0.45\textwidth]{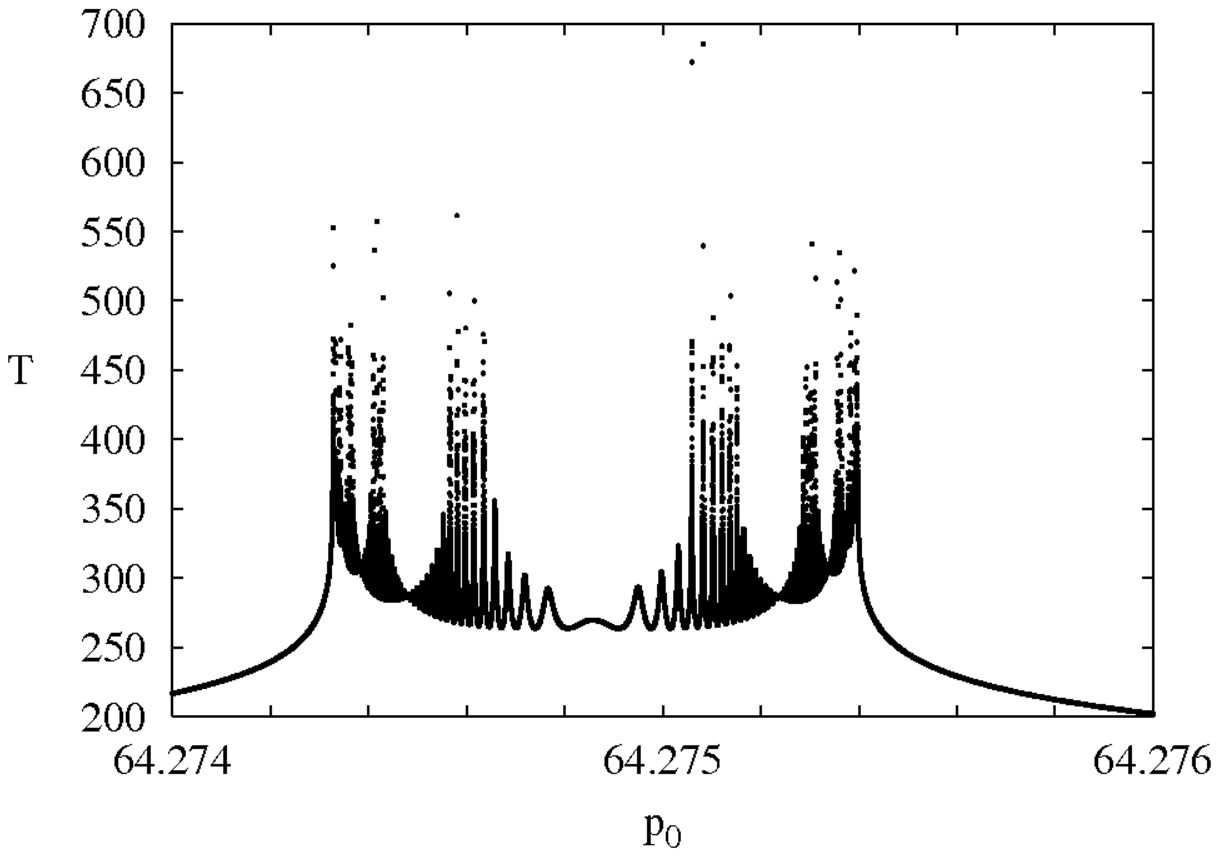}}
\end{center}
\caption{Fock atomic fractal with different resolutions.}
\label{fig9}
\end{figure}
Fig.~\ref{fig9} shows
the function $T(p_0)$ with the normalized detuning $\delta=0.4$, the
recoil frequency $\alpha=10^{-3}$, and the initial number of
cavity photons $n = 10$.
The exit-time function demonstrates an intermittency of smooth
curves and complicated structures that cannot be resolved in principle, no
matter how large the magnification factor.
Fig.~\ref{fig9}b shows magnification of the function for the small interval
$64.1\leqslant p_0\leqslant 64.6$. Further magnification in the range
$64.2743\leqslant p_0\leqslant 64.2754$ shown in Fig.~\ref{fig9}c  reveals
a beautiful self-similar structure.
Some structures in Fig.~\ref{fig9}a that look like fractal are not, in fact,
unresolvable and self-similar. Magnification of the structure in the range
$73.2\leqslant p_0\leqslant 73.8$ demonstrates quite a smooth function without unresolvable
substructures and with only two singular points on the borders of the
respective momentum interval.
Beating  in all the structures
of the atomic fractal in Fig.~\ref{fig9} should be attributed
to the structure of the Hamilton-Schr\"odinger equations (\ref{29}) which
describe two atom-field oscillators with slightly different frequencies.

The exit time
$T$, corresponding to both smooth and unresolved $p_0$ intervals, increases
in average with increasing the magnification factor. It follows that there exist atoms
never reaching the detectors in spite of the fact that they have no obvious
energy restrictions to leave the cavity. Tiny interplay between chaotic external
and internal dynamics prevents these atoms from leaving the cavity. The similar
phenomenon in Hamiltonian systems is known as {\it dynamical trapping}
\cite{Z98}. Different
kinds of atomic trajectories, which are computed with the system (\ref{29}),
are shown in Fig.~\ref{fig8}b. A trajectory with the number $m$ transverses
the central node of the standing-wave, before being detected, $m$ times and is
called $m$-th trajectory. There are also special separatrix-like
$mS$-trajectories following which atoms in infinite time reach the
stationary points $x_s=\pm\pi s$ ($s=0,\, 1,\, 2,\,\dots $),
$p_s=0$, transversing $m$ times the central node. These points are
the anti-nodes of the standing wave where the force acting on atoms is zero.
A detuned atom can
asymptotically reach one of the stationary points after transversing the
central node $m$ times.
The trajectory with number 1, showing in Fig.~\ref{fig8}b, is close to a
separatrix-like $1S$-trajectory.
The smooth $p_0$ intervals in the first-order
structure in Fig.~\ref{fig9}a correspond to atoms transversing
once the central node and reaching the right detector. The unresolved singular
points in the first-order structure with $T=\infty$ at the border between the
smooth and unresolved $p_0$ intervals are generated by the
$1S$-trajectories. Analogously, the smooth $p_0$ intervals
in the second-order structure in Fig.~\ref{fig9}b correspond to
the 2-nd order trajectories with singular points between them
corresponding to the $2S$-trajectories and so on.

There are two different mechanisms of generation of infinite exit times,
namely,
dynamical trapping with infinite oscillations ($m=\infty$) in a cavity and the
separatrix-like motion ($m\ne\infty$). The set of all initial momenta generating
the separatrix-like trajectories is a countable fractal. Each point in the set
can be specified as a vector in a Hilbert space with $m$ integer nonzero
components. One is able to prescribe to any unresolved interval of a 
$m$-th order structure a set with $m$ integers, where the first integer is a
number of a second-order structure to which trajectory under consideration
belongs in the first-order structure, the second integer is a number of a
third-order structure in the second-order structure mentioned above, and so on.
Such a number set is analogous to a directory tree address:
``$<$a subdirectory of the root directory$>$/$<$a subdirectory of the 2-nd
level$>$/$<$a subdirectory of the 3-rd level$>$/...''.
Unlike the separatrix fractal, the set
of all initial atomic momenta leading to dynamically trapped atoms with
$m=\infty$ seems to be uncountable.
\begin{figure}[ht]
\begin{center}
{\bf (a)} \parbox[t]{0.45\textwidth}{\rule{0pt}{0pt}\\
\begin{center}
\includegraphics[width=0.45\textwidth]{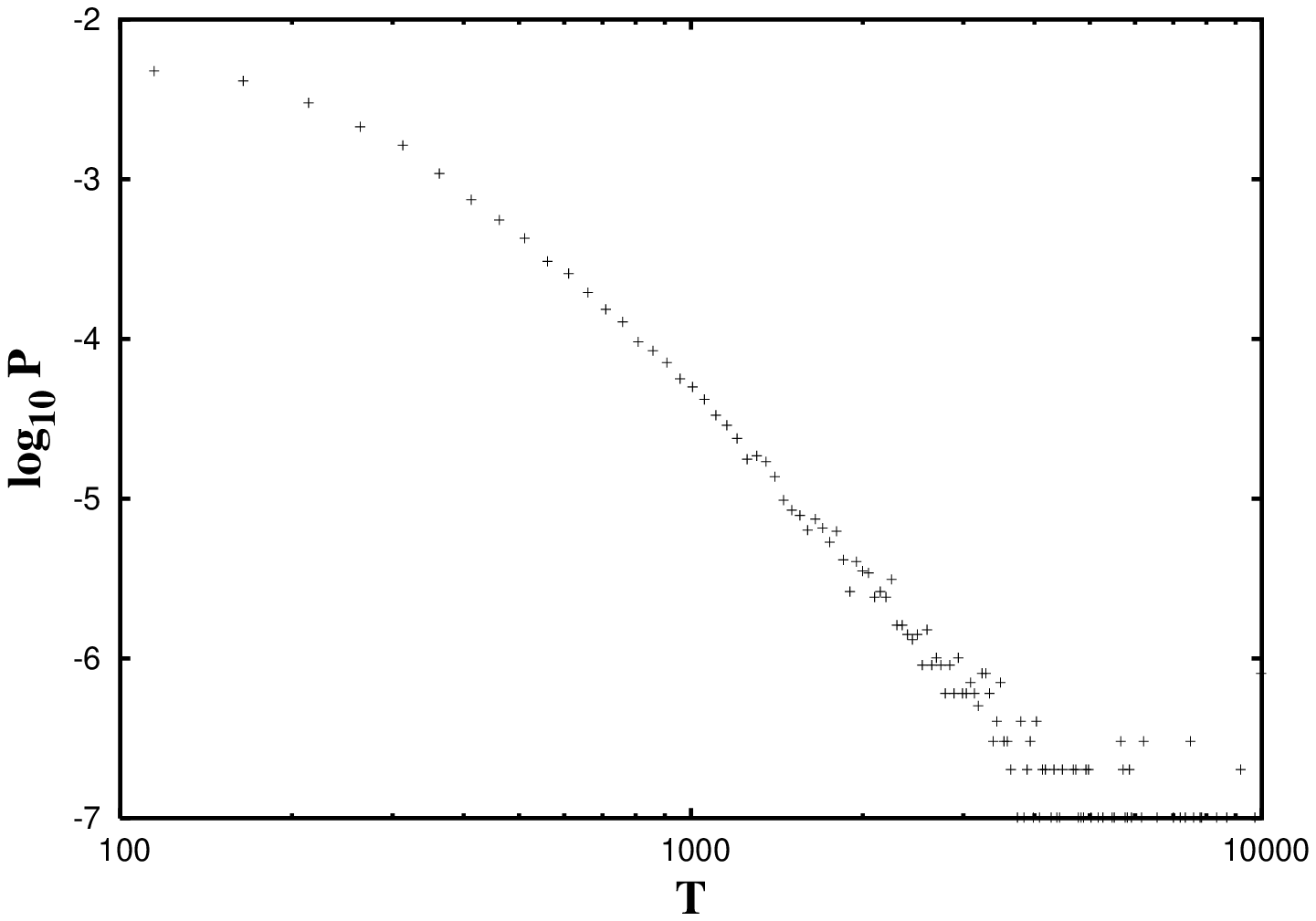}
\end{center}}\\
{\bf (b)} \parbox[t]{0.45\textwidth}{\rule{0pt}{0pt}\\
\includegraphics[width=0.45\textwidth]{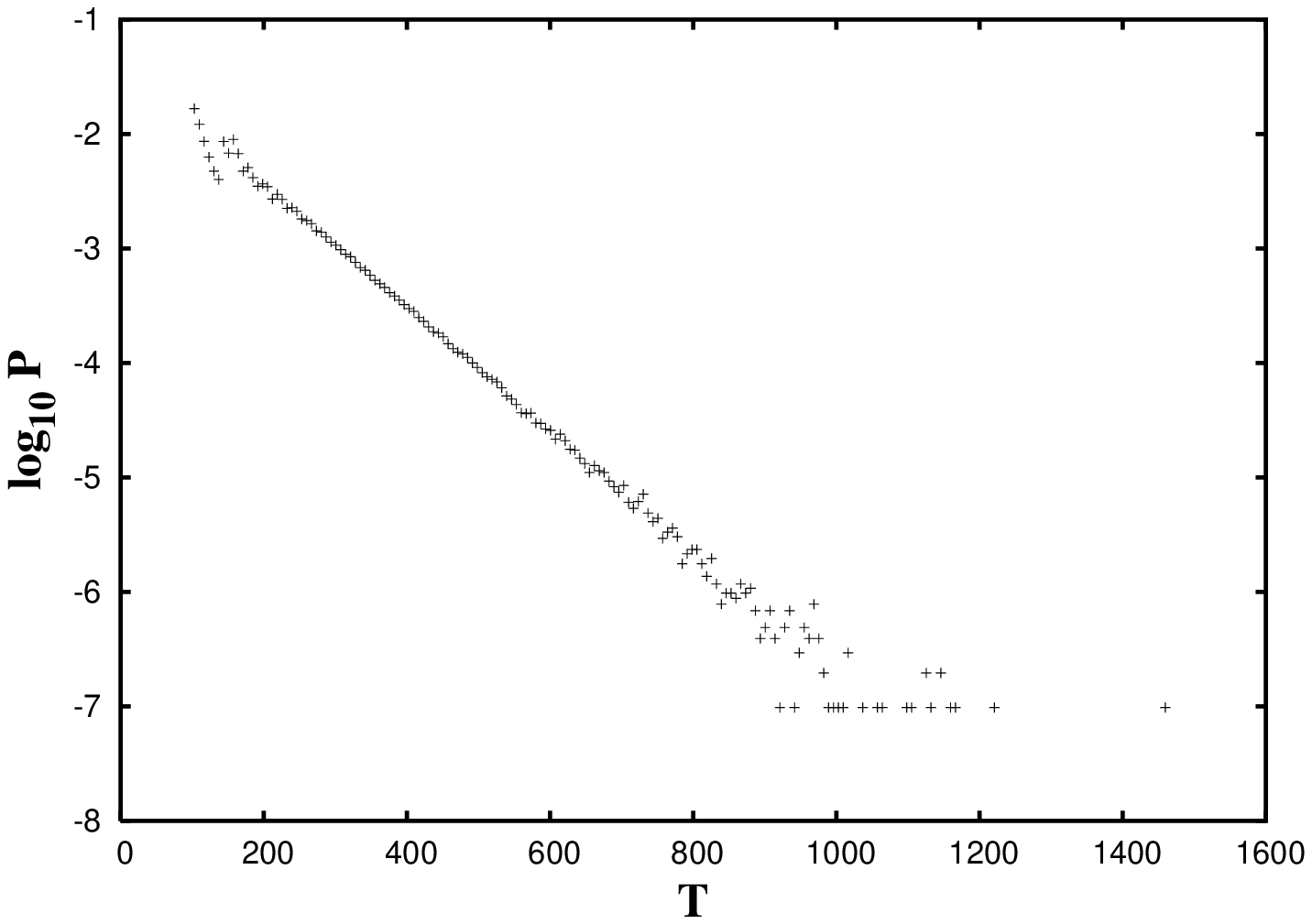}}
\end{center}
\caption{Exit-time distributions in the Fock field: (a) an algebraic decay 
in the range of initial atomic momenta $8\le p_0\le40$ and an exponential 
decay in the range of initial atomic momenta $40\le p_0\le41$.}
\label{fig10}
\end{figure}

Nonlinear  fractal dynamics implies specific statistical properties 
of chaotic motion in Hamiltonian systems (for a review see \cite{Z98}). 
Even a finite-dimensional atom-field system's phase space has a complicated 
topology whose simplified image is given by projections of Poincar\'e 
sections and  fractal scattering functions. 
Tiny interplay between all the degrees of freedom is responsible for 
dynamical trapping of atoms even in a very short microcavity and for anomalous 
statistical properties of the fundamental atom-photon interaction.  
The probability distribution of exit times $P(T)$ for 
$2 \cdot 10^5$ events with initial atomic momenta in the range 
$8\le p_0\le40$  is shown in 
Fig.~\ref{fig10}a in double logarithmic scale. It is close 
to a Poissonian distribution with comparatively short exit times up to $T \simeq 
300$ and demonstrates an algebraic decay with the characteristic exponent 
$\gamma \simeq -3.72$ at its tail ($ \le 300 T \le 4000$).  In the 
range of initial atomic momenta $40\le p_0\le41$, the respective PDF 
for $1.5 \cdot 10^6$ events is computed under the same other conditions 
to be a Poissonian-like. 

\subsection{Coherent fractal}

\begin{figure}[ht]
\begin{center}
{\bf (a)} \parbox[t]{0.45\textwidth}{\rule{0pt}{0pt}\\
\includegraphics[width=0.45\textwidth]{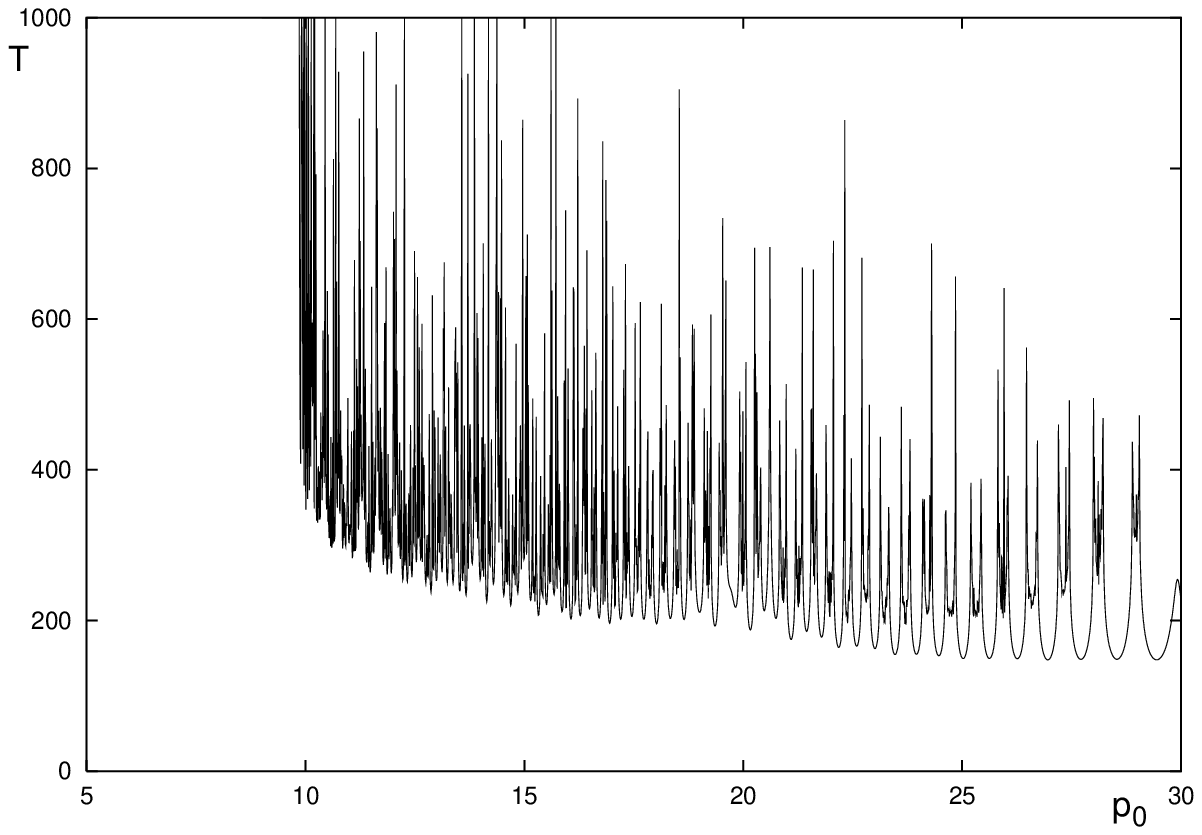}}\\
{\bf (b)} \parbox[t]{0.45\textwidth}{\rule{0pt}{0pt}\\
\includegraphics[width=0.45\textwidth]{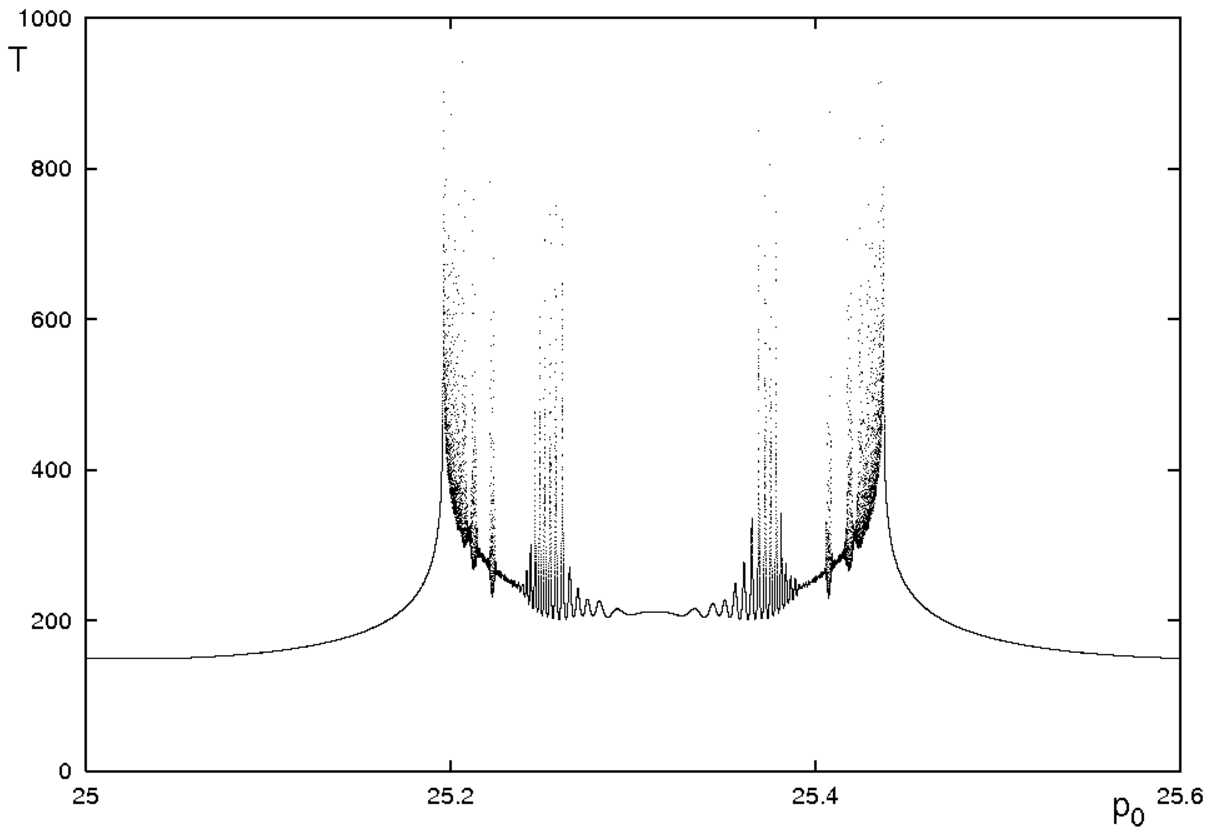}}\\
{\bf (c)} \parbox[t]{0.45\textwidth}{\rule{0pt}{0pt}\\
\includegraphics[width=0.45\textwidth]{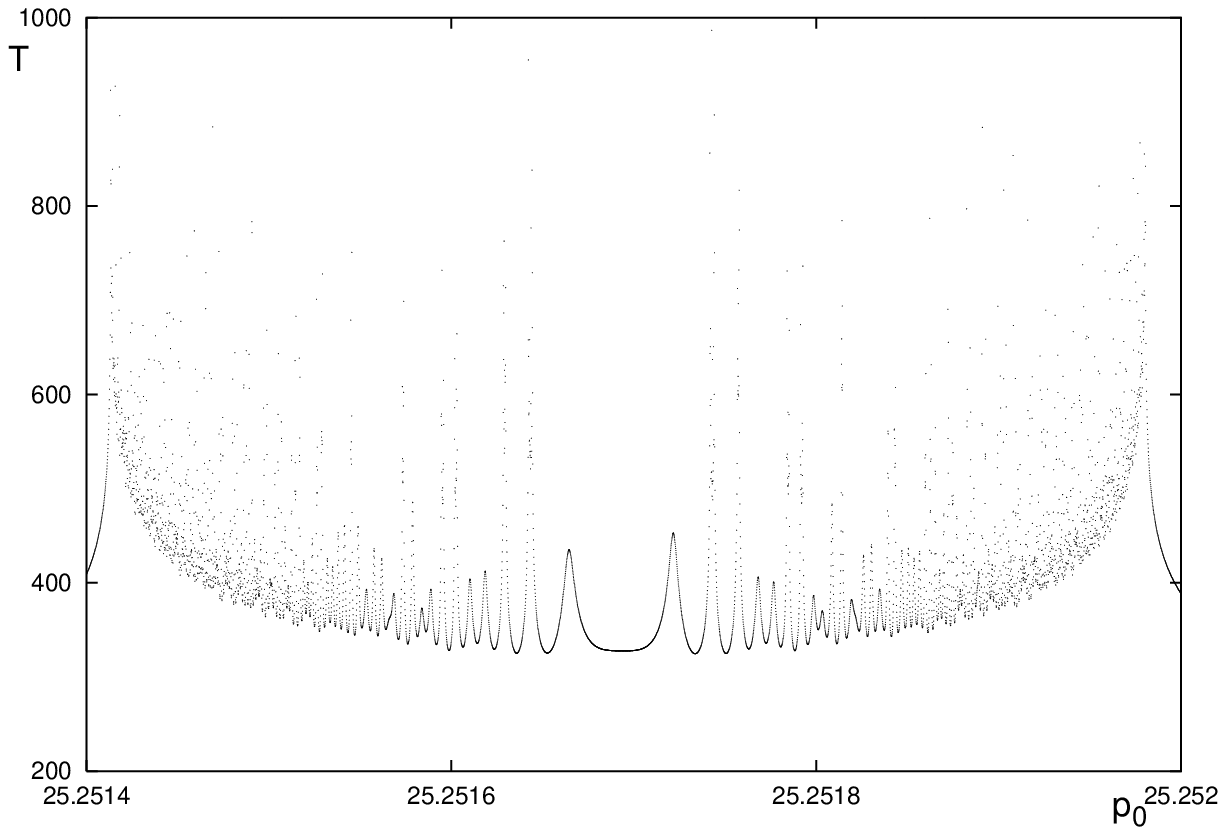}}
\end{center}
\caption{Coherent atomic fractal with different resolutions.}
\label{fig11}
\end{figure}

Following to the scheme in Fig.~\ref{fig8}a, let us consider in this section 
scattering of atoms by the standing wave field initially prepared in the 
coherent state (\ref{19}). 
Before injecting into a cavity, atoms supposed to 
be prepared in the excited state $\left|2\right>$ with $z(0)=1$. In this case 
the initial  conditions  are
\begin{equation}
\begin{aligned}
u_n(0)&=v_n(0)=0,\quad \forall n,\\ 
z_n(0)&=e^{-<n>}\frac{<n>^n}{n!}.
\end{aligned}
\label{33}
\end{equation}
Fig.~\ref{fig11} demonstrates the respective exit-time function $T(p_0)$ 
with $\delta=0.1$, $\alpha=0.001$, and $<n> = 10$. 
A coherent state is not a sharply defined state, as a Fock one, but
a superposition of an infinite number of Fock states. We used a truncated basis 
of $1000$ Fock states for the cavity mode in our simulation. Resolution of 
one of the unresolved structures in Fig.~\ref{fig11}a is shown in 
Fig.~\ref{fig11}b. It resembles the respective fragment of the Fock fractal 
in Fig.~\ref{fig9}b. Further magnification of the function in Fig.~\ref{fig11}b, 
which 
is shown in Fig.~\ref{fig11}c, reveals again a self-similar structure with 
smooth and singular zones. 

\begin{figure}[ht]
\begin{center}
\parbox[t]{0.45\textwidth}{\rule{0pt}{0pt}\\
\includegraphics[width=0.45\textwidth]{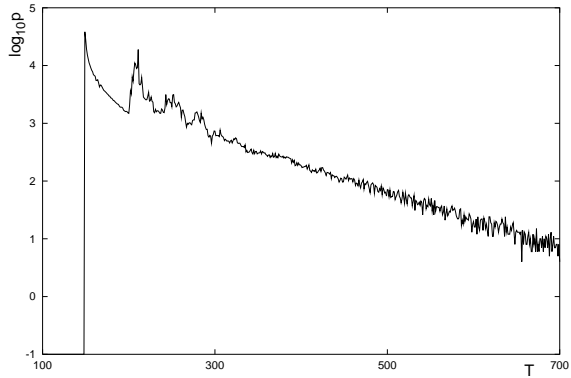}}
\end{center}
\caption{Exit-time distribution in the coherent field.}
\label{fig12}
\end{figure}
        
We collect an exit time statistics with $6 \cdot 10^5$ events by counting atoms with initial momenta 
in the range $9\le p_0\le30$ reaching the detectors. The plot of the respective histogram of
exit times, shown in Fig.~\ref{fig12}, demonstrates a few local maxima. 
The corresponding times
of exit of atoms can be approximately estimated with the help of the formula
(\ref{33}) for resonant atoms. The first maximum around $T\simeq150\div160$ 
corresponds, mainly,
to atoms which, being initially directed to the right with comparatively low
momenta $p_0\simeq10$, turn back before reaching the central node and are 
registered by
the left detector. It follows from (\ref{33}) that for such atoms 
$T_1\simeq\pi/2\alpha p_0\simeq157$ at $\alpha p_0=0.01$. 
However,
the atoms, to be injected initially with the momentum $p_0\simeq20\div25$ 
and registered by the right detector, may contribute to the first maximum 
as well because, after transversing the central node, they can be accelerated 
and gain the values of the momentum $p$ up to $50$. The second local maximum 
around $T\simeq200$ corresponds, mainly, to atoms with $p_0\simeq20\div30$ 
which transverse the central node and are registered by the right detector 
for the time $T_2\sim 3\pi/2\alpha p_0$. The other local maxima of the 
PDF are not so pronounced as the first ones,
they are formed by the atoms transversing the central node a few times. The
PDF in Fig.~\ref{fig12} demonstrates an exponential decay at the tail up to 
the exit times $T = 700$. 

\subsection{Bose-Einstein  fractal}

In conclusion of this section we present in Fig.~13 the fractal 
computed with the  field initially prepared in a Bose-Einstein
state (\ref{20}) and atoms supposed to 
be prepared in the excited state $\left|2\right>$ with $z(0)=1$. 
The initial  conditions  are
\begin{equation}
\begin{aligned}
u_n(0)&=v_n(0)=0,\quad \forall n,\\ 
z_n(0)&= \frac{<n>^n}{(1+<n>)^{n+1}}.
\end{aligned}
\label{34}
\end{equation}
The values of the control parameters are the same as for the coherent fractal. 
Fig.~13 demonstrates again a self-similar structure of the exit-time 
distribution. We want to stress that the Hamilton-Schr\"odinger sets 
of equations generating 
the Fock and the other fractals differ strongly in the number of equations. 
Nevertheless, the respective $T(p_0)$ functions are rather similar in their main 
features. The atomic fractals we have found are generated by dynamical chaos 
in the atom-photon interaction but not by noise.

\begin{figure}[ht]
\begin{center}
{\bf (a)} \parbox[t]{0.45\textwidth}{\rule{0pt}{0pt}\\
\includegraphics[width=0.45\textwidth]{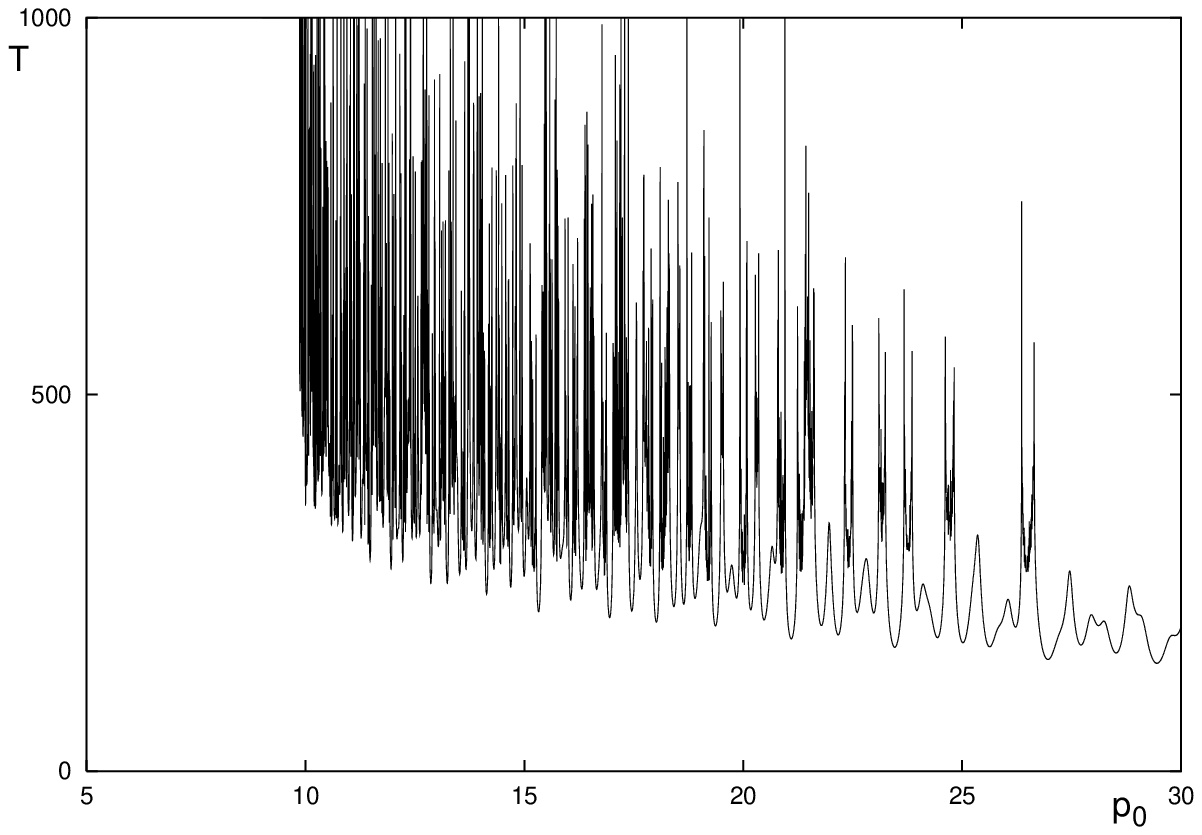}}\\
{\bf (b)} \parbox[t]{0.45\textwidth}{\rule{0pt}{0pt}\\
\includegraphics[width=0.45\textwidth]{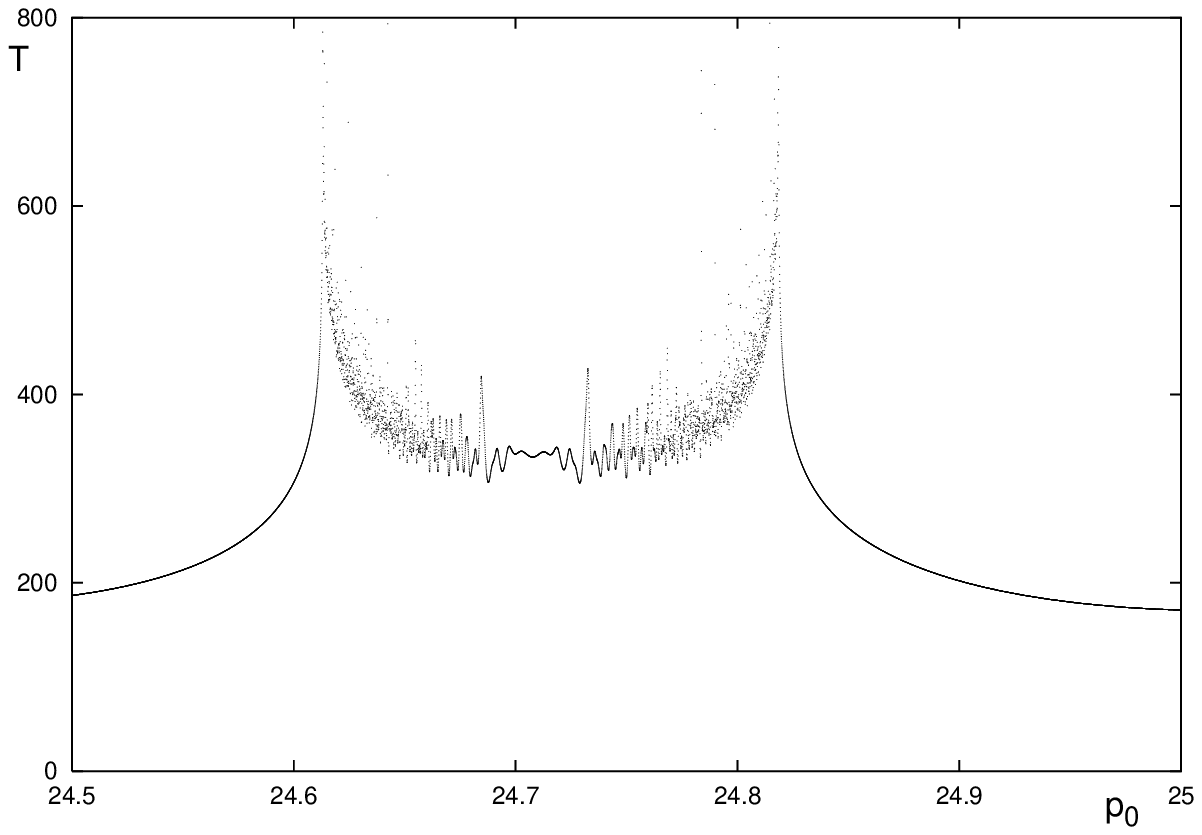}}\\
{\bf (c)} \parbox[t]{0.45\textwidth}{\rule{0pt}{0pt}\\
\includegraphics[width=0.45\textwidth]{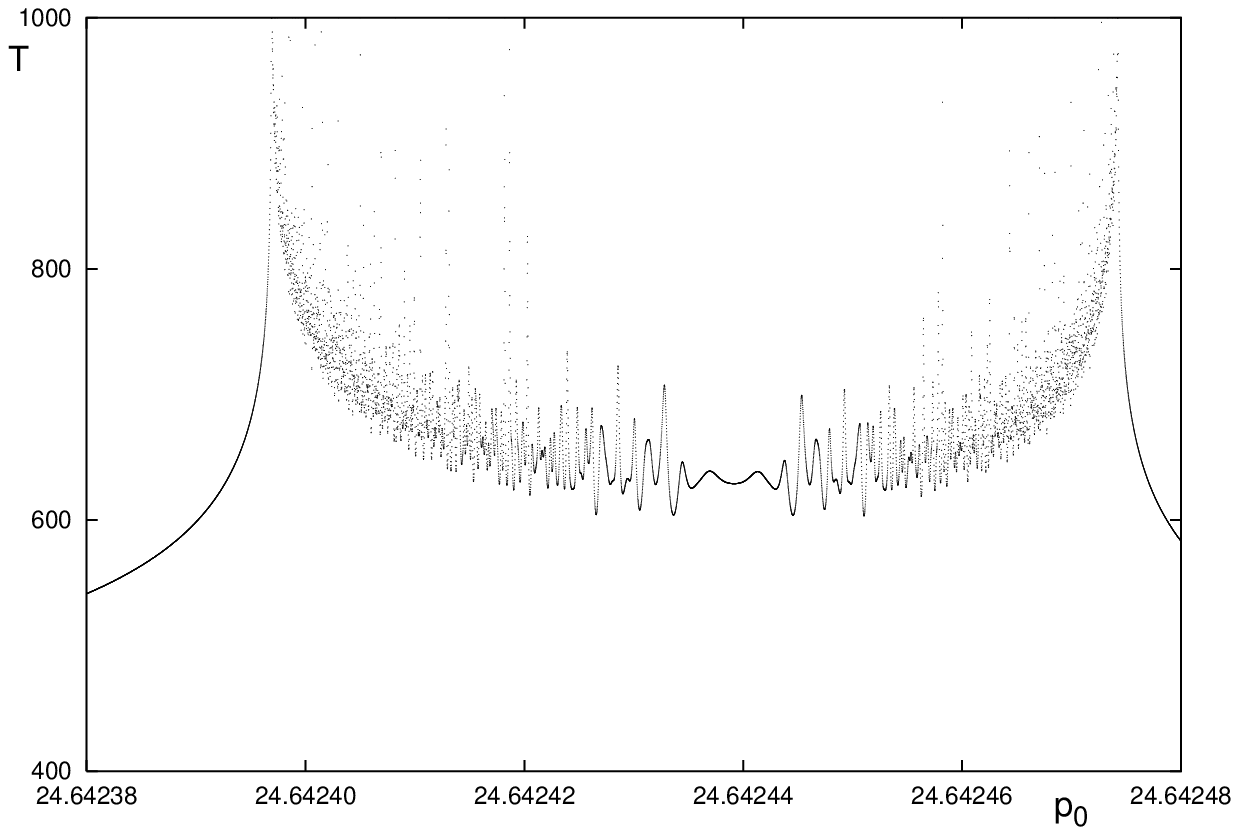}}\\
\end{center}
\parbox[t]{0.48\textwidth}{\rule{0pt}{0pt}
\caption{Bose-Einstein  atomic fractal with different resolutions.}}
\label{fig13}
\end{figure}

\section{Conclusion}

We have studied in the strong-coupling regime the fundamental interaction 
between a two-level atom with recoil and a quantized radiation field in 
a single-mode cavity in the mixed quantum-classical formalism modelling 
quantum evolution of the electronic-field purely quantum system to be disturbed 
by translational atomic motion. We have managed to derive an infinite set of 
the Hamilton-Schr\"odinger nonlinear ODE's for real valued electronic-field 
probability amplitudes and expectation values of the atomic center-of-mass position 
and momentum. These equations of motion are able to demonstrate known features 
of the atom-field quantum evolution in the strong-coupling limit such as 
collapses and revivals of the atomic population inversion modulated by 
atomic motion through the nodes and antinodes of the standing wave. 
We have shown that even in the absence of any other interaction with 
environment the Hamiltion-Schr\"odinger dynamical system provides the emergence  
of classical Hamiltonian dynamical chaos from cavity quantum electrodynamics. 

We have investigated the atom-photon nonlinear dynamics with Fock, coherent and 
Bose-Einstein  quantum states of the initial cavity field. Positive values of the 
maximal Lyapunov exponent $\lambda$ have been found with reasonable values of 
the control parameters, the detuning of the atom-field resonance, $\delta$, 
the atomic recoil frequency, $\alpha$, and the initial mean number of photons. 
Exponential sensitivity to initial conditions may manifest itself in the 
chaotic dependence of the output values of the atomic population inversion on 
its input values that may, in principle, be measured in real experiments. 
New manifestations of Hamiltonian chaos in the atom-photon interaction, we have 
found numerically in the dependence of the atomic exit times on the initial 
momentum, are atomic fractals. They were classifield  by the names of 
the respective quantum field states, i.e., the Fock, coherent and Bose-Einstein   
fractals. We have found anomalous statistical properties of the chaotic 
atom-photon interaction: L\'evy atomic flights through a cavity and a power law  
at the tail of the exit-time distribution function in the Fock field. Power-law 
tails should appear (at least, in some ranges of the control parameters) in 
the PDF's with coherent and Bose-Einstein  fields, as well, but their detection 
would require a longer computation time than we have used.  

\begin{acknowledgments}
I am grateful to Leonid Kon'kov and Michael Uleysky for preparing the figures. 
This work was supported by the Russian Foundation for Basic Research under    
Grant No. 02--02--17796, by the Program ``Mathematical Methods in  
Nonlinear Dynamics" 
of the Russian Academy of Sciences, and by the Program for 
Basic Research of the Far Eastern Division of the Russian Academy of 
Sciences.                
\end{acknowledgments}

\end{document}